\documentclass[twocolumn]{aastex631}

\shorttitle{Instabilities in compact multiplanet systems}
\shortauthors{Lammers et al.}

\graphicspath{{./}{}}
\usepackage{amsmath}
\usepackage{comment}

\begin{document}

\newcommand{\paren}[1]{\left(#1\right)}
\newcommand{\pd}[2]{\frac{\partial #1}{\partial #2}}
\newcommand{\bigO}[1]{\mathcal{O}\paren{#1}}
\newcommand{\celmech}{\texttt{celmech}~}
\newcommand{\fracbrac}[2]{\paren{\frac{#1}{#2}}}

\title{The instability mechanism of compact multiplanet systems}


\author[0000-0001-9985-0643]{Caleb Lammers}
\affiliation{Canadian Institute for Theoretical Astrophysics, University of Toronto, 60 St.\ George Street, Toronto, ON M5S 3H8, Canada}
\affiliation{Department of Physics, University of Toronto, 60 St.\ George Street, Toronto, ON M5S 1A7, Canada}
\affiliation{Department of Astrophysical Sciences, Princeton University, 4 Ivy Lane, Princeton, NJ 08544, USA}

\author[0000-0002-1032-0783]{Sam Hadden}
\affiliation{Canadian Institute for Theoretical Astrophysics, University of Toronto, 60 St.\ George Street, Toronto, ON M5S 3H8, Canada}

\author[0000-0002-8659-3729]{Norman Murray}
\affiliation{Canadian Institute for Theoretical Astrophysics, University of Toronto, 60 St.\ George Street, Toronto, ON M5S 3H8, Canada}
\affiliation{Department of Physics, University of Toronto, 60 St.\ George Street, Toronto, ON M5S 1A7, Canada}

\begin{abstract}

To improve our understanding of orbital instabilities in compact planetary systems, we compare suites of $N$-body simulations against numerical integrations of simplified dynamical models. We show that, surprisingly, dynamical models that account for small sets of resonant interactions between the planets can accurately recover $N$-body instability times. This points toward a simple physical picture in which a handful of three-body resonances, generated by interactions between nearby two-body mean motion resonances, overlap and drive chaotic diffusion, leading to instability. Motivated by this, we show that instability times are well described by a power law relating instability time to planet separations, measured in units of fractional semi-major axis difference divided by the planet-to-star mass ratio to the $1/4$ power, rather than the frequently adopted $1/3$ power implied by measuring separations in units of mutual Hill radii. For idealized systems, the parameters of this power-law relationship depend only on the ratio of the planets’ orbital eccentricities to the orbit-crossing value, and we report an empirical fit to enable quick instability time predictions. This relationship predicts that observed systems comprised of three or more sub-Neptune-mass planets must be spaced with period ratios $\mathcal{P}\,{\gtrsim}\,1.35$ and that tightly spaced systems ($\mathcal{P}\,{\lesssim}\,1.5$) must possess very low eccentricities ($e\,{\lesssim}\,0.05$) to be stable for more than $10^9$ orbits.

\end{abstract}


\section{Introduction}
\label{sec:intro}

NASA's \emph{Kepler} mission revealed that planetary systems comprised of multiple sub-Neptune-sized planets orbiting in compact configurations on short-period orbits ($P\,{\sim}\,10$\,days) are commonplace around sun-like stars \citep[e.g.,][]{Borucki2011, Lissauer2011, Fressin2013, Fabrycky2014, Zhu2018}. The orbital architectures of these systems, markedly distinct from the solar system's, have inspired numerous investigations into their long-term dynamical evolution. A key finding of these studies is that many transiting systems appear perched on the verge of instability, with expected dynamical lifetimes similar to their present ages and no ability to stably accommodate additional planets \citep{Fang&Margot2013, Pu&Wu2015, Volk&Gladman2015, Obertas2023}.

The fact that many transiting sub-Neptune systems are perched on the verge of instability is interpreted as evidence that dynamical instabilities in their past have played a central role in setting the orbital configurations we find them in today. In this picture, systems emerge from the protoplanetary disk in dynamically unstable configurations. Instabilities then lead to collisions and mergers\footnote{In \emph{Kepler} multiplanet systems, the ratios of surface escape velocities to orbital velocities (i.e., the Safronov numbers) are such that dynamical instabilities should almost always culminate in collisions rather than ejections of planets from the system.} that leave behind remade systems with fewer planets on wider orbital spacings. In these new orbital configurations, the timescale over which chaos and dynamical instabilities manifest is increased. Given enough time, instabilities eventually arise again, producing even more widely spaced systems with fewer planets. In this way, systems are thought to evolve in a state of ``perpetual meta-stability" such that they will typically exhibit dynamical lifetimes similar to their current age \citep[e.g.,][]{Laskar1996}. Support for this picture is further bolstered by recent work \citep{Poon2020, Lammers2023, Ghosh&Chatterjee2024} which has shown that giant impacts can reproduce several of the trends seen in the observed exoplanet population.

While the available evidence for the importance of dynamical instabilities in shaping planetary systems is compelling, our incomplete theoretical understanding of dynamical chaos and instabilities in these systems is an impediment to drawing more definitive conclusions. Thus, advancing our understanding of the long-term dynamics of these systems is critical for developing a more comprehensive theory of their formation and evolution. The goal of this paper is to clarify the underlying dynamical mechanism leading to instabilities in compact multiplanet systems. We do so by developing simplified dynamical models for the gravitational interactions in multiplanet systems and evaluate them by comparing their predictions against the results of $N$-body simulations. Before presenting our results, we briefly review what $N$-body simulations have revealed about the phenomenology of instabilities in compact systems (§~\ref{sec:prevsims}) and discuss what is known from analytical investigations (§~\ref{sec:prevtheory}).

\subsection{Previous work: numerical experiments}
\label{sec:prevsims}

Lacking a comprehensive theoretical understanding of the planetary $N$-body problem, many authors have turned to numerical experiments to investigate dynamical instabilities in multiplanet systems. These experiments revealed that systems comprised of three or more planets can exhibit chaos and instability on a broad range of timescales. Early numerical investigations of coplanar, equally spaced, initially circular systems found that instability times depend exponentially on initial planet spacing, measured in units of the planets' mutual Hill radii, with a fairly weak dependence on planet mass and the number of planets \citep{Chambers1996}. Numerous subsequent studies have derived empirical relationships between instability times and initial planet spacing, measured in mutual Hill radii, that reproduce this same basic result: Instability times increase approximately exponentially as systems comprised of multiple planets become more widely spaced \citep[e.g.,][]{Yoshinaga1999, Zhou2007, FaberQuillen2007, Smith&Lissauer2009, Funk2010, Pu&Wu2015, Morrison&Kratter2016, Obertas2017, Gratia&Lissauer2021, Lissauer&Gavino2021}.

Most numerical studies have formulated empirical relationships by measuring planet separations in terms of mutual Hill radii, which is proportional to $\fracbrac{m}{M_*}^{1/3}$, where $\fracbrac{m}{M_*}$ is the planet-to-star mass ratio. However, it was noted already by \citet{Chambers1996} that normalizing planet separations by a factor of $\fracbrac{m}{M_*}^{1/4}$, rather than $\fracbrac{m}{M_*}^{1/3}$, leads to empirical criteria that better predict instability times over a wide range of masses. As we will see below, this $\fracbrac{m}{M_*}^{1/4}$ scaling is expected on theoretical grounds \citep[see also][]{Yalinewich&Petrovich2020}.

\citet{Obertas2017} integrated systems comprised of five equally spaced, Earth-mass planets on coplanar and initially circular orbits around a solar-mass star. In addition to the previously noted exponential increases in instability times with planet spacing, their simulations revealed that survival times dip sharply at spacings corresponding to first- and second-order mean motion resonances (MMRs) between adjacent planets. \citet{Gratia&Lissauer2021} observed similar, but less significant, dips in survival time in the vicinity of low-order MMRs when simulating systems comprised of initially eccentric planets. Theoretical works have identified the role of low-order MMRs in causing dynamical chaos in compact planetary systems, and our results below will underscore their centrality to producing the instabilities observed in $N$-body simulations.

\subsection{Previous work: theory}
\label{sec:prevtheory}
 
The theoretical account of the origins of dynamical chaos in systems comprised of two closely spaced, coplanar planets is relatively complete. This theory is built on the concept of resonance overlap \citep{Walker&Ford1969, Chirikov1979}, which predicts that dynamical chaos in conservative systems occurs in regions of the phase space where two or more nonlinear resonances are predicted to occur when considering each individual resonance in isolation. \citet{Wisdom1980} was the first to apply the resonance overlap criterion to this problem, deriving a criterion for the onset of chaos in the circular restricted three-body problem based on the overlap of first-order MMRs. \citet{Wisdom1980}'s result was later generalized by \citet{Deck2013}, who derived a criterion for the critical spacing of a pair of massive planets on nearly circular orbits. Later, \citet{Hadden2018} derived a general criterion for the onset of chaos in two-planet systems without restrictions on their masses or eccentricities. To deal with the infinite number of higher-order MMRs that possess nonzero width in eccentric systems, they showed that the onset of chaos could be predicted by computing the local filling fraction of these MMRs in a region of phase space.

\citet{Petit2020} also produced a theoretical account of chaos and instabilities in systems comprised of three coplanar and initially circular planets. By building on a theory initially put forth by \citet{Quillen2011} \citep[see also][]{Quillen2014}, they were able to successfully predict when the overlap of zeroth-order, three-body resonances (3BRs) produces chaotic behavior. This overlap criterion relies on a similar calculation of local resonance filling fraction to handle the infinite number of 3BRs. Furthermore, \citet{Petit2020} developed a theory of the chaotic diffusion produced by these overlapped resonances that captures the scaling of instability time with the initial spacings and masses of the planets.

\citet{Tamayo2021} explored how, in multiplanet systems, secular evolution of the planets' eccentricities and apsidal alignments can evolve adjacent planet pairs into the regime of resonance overlap described by \citet{Hadden2018}. They also accounted for the enhancement of the local filling fraction of two-body MMRs that results from introducing additional planets. In other words, they adapted the MMR overlap criterion of \citet{Hadden2018} to include the possibility that the MMRs doing the overlapping involve more than just one pair of planets. \citet{Tamayo2021}'s account of chaos in general, noncircular multiplanet systems, based on MMR overlap, is apparently at odds with the results of \citet{Petit2020}, who find 3BR overlap responsible for chaos in initially circular systems.

\citet{Rath2022} developed a general theory of the origins of chaos in three-planet systems that resolves this apparent contradiction. In particular, they showed that the usual resonance overlap criterion is only a crude guide to the extent of chaos generated by the overlap of MMRs involving distinct planet pairs. In general, the chaos generated locally by the interaction of two such MMRs extends beyond the phase-space volume occupied by the resonances themselves. \citet{Rath2022} demonstrated that this is because the resonant interactions generate a family of 3BRs and show that the extent of chaos can be predicted based on where this family of 3BRs overlaps.

While the theory of \citet{Rath2022} predicts when chaos will occur in generic three-planet systems, it does not address the timescales on which dynamical instabilities arising from this chaos are expected to occur. Additionally, the chaotic diffusion theory developed by \citet{Petit2020} does not readily generalize to the case of initially eccentric systems --- special symmetries exist in the circular case, restricting chaotic diffusion to a single direction in phase space, that are no longer present when considering eccentric systems. Thus, our theoretical understanding of dynamical instabilities in multiplanet systems remains incomplete.

Finally, we note that the use of computer algebra to construct and study simplified dynamical models has proven fruitful for understanding the chaotic behavior generated by interactions among secular resonances in the inner solar system \citep[e.g.,][]{Mogavero&Laskar2021, Hoang2022, MogaveroLaskar2022}. In this paper, we adopt a similar computer-algebra-assisted approach to investigate dynamical chaos and instabilities arising from the interactions of resonances in tightly spaced planetary systems.

\subsection{This paper}

The goal of this paper is to advance our theoretical understanding of the underlying mechanism driving instabilities in compact multiplanet systems. We do so by comparing ensembles of $N$-body simulations against simplified dynamical models that account for limited sets of resonant interactions between the planets in the simulated systems. Using the open-source \celmech code \citep{celmech2022}, we are able to efficiently generate and integrate equations of motion of different dynamical models that account for various combinations of resonant interactions.

The layout of the paper is as follows. In Section~\ref{sec:simulations} we describe our numerical simulations, including both $N$-body simulations (§~\ref{sec:simsetup}) and our simplified dynamical models (§~\ref{sec:Hamtheory}). Our $N$-body results are presented in Section~\ref{sec:results}, and we compare results from our dynamical models and $N$-body simulations in Section~\ref{sec:Hamiltonian_models}. We discuss our results in Section~\ref{sec:discussion} and conclude in Section~\ref{sec:conclusion}.

\section{Simulation setups}
\label{sec:simulations}

\subsection{$N$-body simulations}
\label{sec:simsetup}
We conduct $N$-body simulations of planetary systems comprised of five coplanar, equally spaced planets orbiting a solar-mass star. Our simulations are carried out with the open-source \texttt{REBOUND} code \citep{Rein2012}. We use the symplectic Wisdom-Holman integrator \texttt{WHFast} \citep{Wisdom&Holman1991, Rein2015} with the time step set to $P_1/20$, where $P_1$ is the initial orbital period of the innermost planet. Simulations are integrated for a total time of $10^7\,P_1$ or stopped once a pair of planets' orbits overlap. Specifically, we stop $N$-body simulations and record $t_\mathrm{inst}$ when $(1\,{-}\,e_{i\,{+}\,1})a_{i\,{+}\,1}\,{-}\,(1\,{+}\,e_i)a_i\,{<}\,d$, where $d\,{=}\,a_1\fracbrac{m}{M_*}^{1/3}$ is the Hill radius of the innermost planet.\footnote{$N$-body studies typically monitor instead for close encounters between planets. This is not a practical stopping condition for the dynamical model integrations presented in this work, because in this case integration time steps are large enough that close encounters can be missed. We have repeated some of our $N$-body simulations with a minimum distance threshold and found no meaningful change in the recorded instability times.}

We divide our simulations into multiple ensembles, each comprised of $15$,$000$ systems. All planets within a given ensemble are assigned the same mass. Individual systems within each ensemble are initialized so that adjacent planet pairs share a common period ratio, $\mathcal{P}$. This common period ratio is drawn uniformly from an interval $[\mathcal{P}_\mathrm{min},\mathcal{P}_\mathrm{max}]$. The upper and lower limits of this interval are chosen via trial-and-error so that instability times range from ${\sim}\,P_1$ to ${\sim}\,10^7P_1$ and therefore depend on both the planet masses and eccentricities.

The planets' orbital eccentricities are assigned based on the period ratio spacing of the system they reside in. Within each ensemble, eccentricities are set to be a constant fraction of the orbit-crossing eccentricity:
\begin{equation}
\label{eq:ecross_def}
    e_\mathrm{cross}(\mathcal{P}) = \frac{\mathcal{P}^{2/3}-1}{\mathcal{P}^{2/3}+1}~.
\end{equation}
This definition of orbit-crossing eccentricity is chosen such that, if both planets in a pair of adjacent planets have period ratio $\mathcal{P}$ and eccentricities $e\,{=}\,e_{\mathrm{cross}}(\mathcal{P})$, their orbits can overlap. \citet{Hadden2018} show that, for closely spaced planets, the functional dependence of the strengths of MMRs of a given order on eccentricity, $e$, can expressed solely in terms of $e/e_\mathrm{cross}$. Thus, by choosing eccentricities to be a constant fraction of the orbit-crossing value, the relative strengths of MMRs of different orders will be roughly the same for all systems within a given ensemble, irrespective of their period ratios. Finally, planets' initial mean longitudes and longitudes of periapsis are randomly drawn from $[0,\,2\pi)$.

We generate ensembles of simulations with three planet masses, $m_i\,{\in}\,\{0.01,\,1.00,\,100\}\,M_\oplus$, and three normalized eccentricities, $e_i/e_\mathrm{cross}\,{\in}\,\{0.00,\,0.25,\,0.50\}$, resulting in a total of nine ensembles. The results of these simulations are presented in Section~\ref{sec:results}.

\subsection{Simplified dynamical models}
\label{sec:Hamtheory}

We complement our direct $N$-body simulations with integrations of equations of motion that account for limited subsets of resonant interactions between the planets. We use the \celmech code \citep{celmech2022} to generate these equations of motion based on a classic disturbing function expansion of planets' gravitational interaction potential. Here, we provide a brief overview of our disturbing-function-based approach for generating and integrating equations of motion. We refer the reader to \citet{celmech2022} for technical details.

The Hamiltonian governing a system of $N$ planets orbiting a common star can be written as
\begin{equation} 
\label{Hamil}
H = \sum_{i=1}^{N} H_{\mathrm{Kep},i} + \sum_{i=1}^{N} \sum_{j=i+1}^{N} H_{\mathrm{int.}}^{(i,j)}
\end{equation}
\noindent
where $H_{\mathrm{Kep},i}$ describes the interaction between the star and the $i$th planet, and 
\begin{equation}
    H_{\mathrm{int.}}^{(i,j)} = -\frac{Gm_im_j}{|\mathbf{r}_i- \mathbf{r}_j|} + \frac{\mathbf{p}_i\cdot\mathbf{p}_j}{M_*}
\end{equation}
where $\mathbf{r}_i$ is the position vector of the $i$th planet and $\mathbf{p}_i$ is its momentum. In terms of heliocentric Keplerian orbital elements $(a_i,\,e_i,\,I_i,\,\lambda_i,\,\varpi_i,\, \Omega_i)$, the Keplerian terms in Eq.~\ref{Hamil} are simply given by $H_{\mathrm{Kep},i}\,{=}\,-\frac{1}{2}{G M_* m_i}/{a_i}$, where $G$ is Newton's gravitational constant, $M_*$ is the mass of the central star, and $m_i$ is the mass of the $i$th planet. While the interaction terms, $H_{\mathrm{int.}}^{(i,j)}$, do not admit simple closed-form expressions in terms of the Keplerian orbital elements, they can be expanded as a power series in orbital eccentricities and inclinations and a cosine series in the angular orbital elements \citep[see, e.g.,][]{Murray&Dermott1999, Tremaine2023}. For coplanar orbits ($I_i\,{=}\,0$) and $a_i\,{<}\,a_j$, the expansion of the interaction terms can be written as
\begin{align} 
\label{Hint}
H_{\mathrm{int.}}^{(i,j)} &= -\frac{G m_i m_j}{a_j} \sum_{k_1, k_2, k_3, k_4} \sum_{\nu_3, \nu_4 = 0}^{\infty} \tilde{C}_{(k_1,k_2,k_3,k_4,0,0)}^{(0,0,\nu_3,\nu_4)}(\alpha_{i,j}) \nonumber\\
&e_i^{|k_3| + 2\nu_3} e_j^{|k_4| + 2\nu_4} \cos(k_1 \lambda_j + k_2 \lambda_i + k_3 \varpi_i + k_4 \varpi_j)
\end{align}
where $\alpha_{i,j}\,{=}\,a_i/a_j$ and $k_1,\,k_2,\,k_3,\,k_4$ and $\nu_3,\,\nu_4$ are integer indices. The rotational symmetry of the problem dictates that the coefficients $\tilde{C}$ are only nonzero when $\sum_l^4\,k_l\,{=}\,0$.

Because planet masses are small relative to the central star's mass, standard methods of Hamiltonian perturbation theory can be applied to derive a near-identity transformation from planets' osculating orbital elements to new ``mean" orbital elements. After this transformation is applied, the new Hamiltonian contains only a handful of cosine terms from the multiply infinite sum in Eq.\eqref{Hint} in the new equations of motion (at lowest order in the planet-to-star mass ratio). These will be the terms with slowly varying cosine arguments, including those with $k_1\,\dot{\lambda}_j + k_2\,\dot{\lambda}_i\,{\sim}\,0$ (``MMR terms'') and $k_1\,{=}\,{k_2}\,{=}\,0$ (``secular terms''), both of which play an important role, in general, of determining the system's long-term dynamics.

\begin{figure*}
\centering
\includegraphics[width=0.65\textwidth]{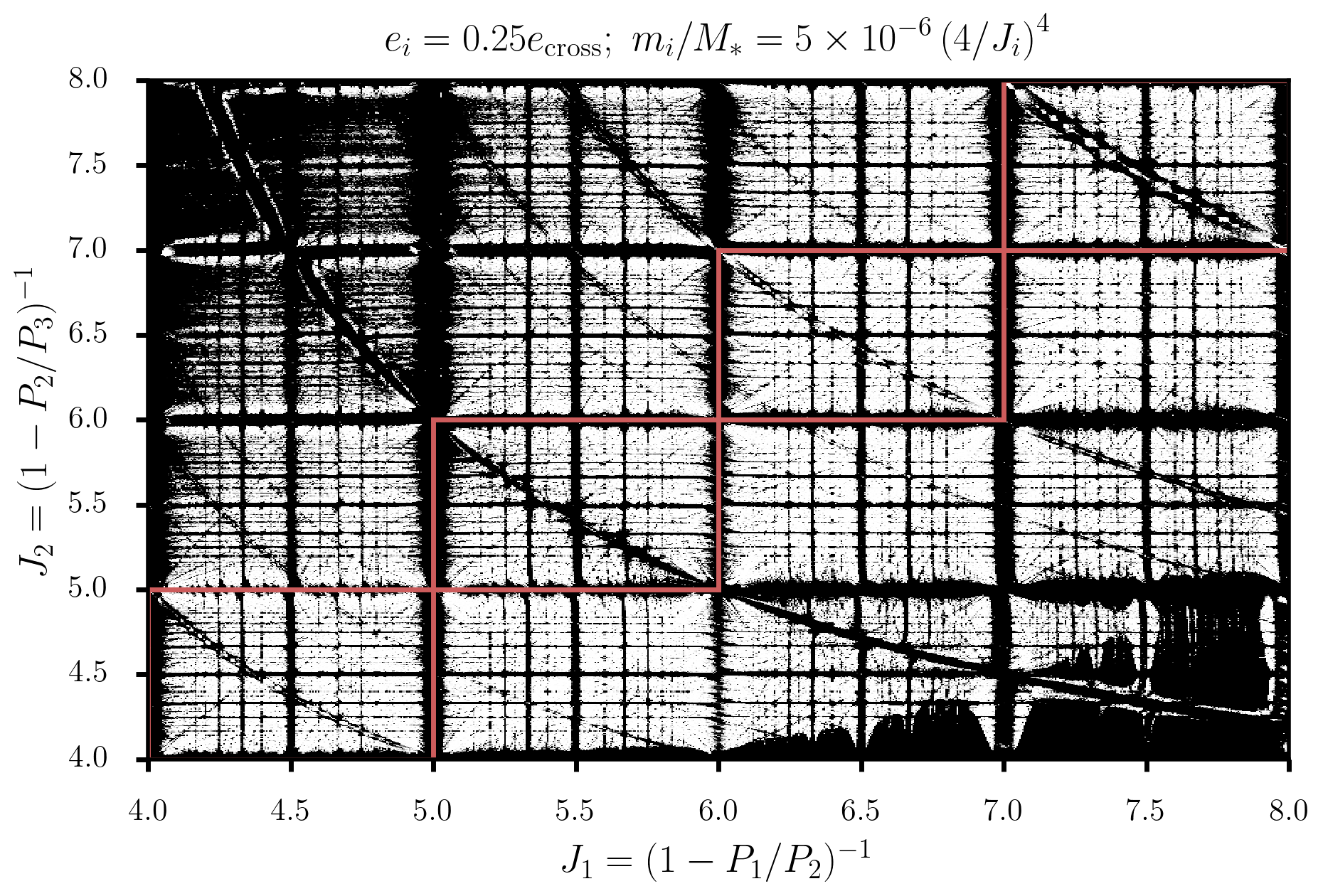}
\caption{Dynamical map highlighting chaotic (black) and regular (white) initial conditions based on the MEGNO chaos indicator, as measured from short $N$-body simulations ($5\,{\times}\,10^4\,P_2$) of a test particle with both inner and outer perturbing planets. The planet masses are scaled with planet spacing ($m_i\,{=}\,5\,{\times}\,10^{-6}\,M_*\fracbrac{4}{J_i}^4$) so that the local density of 3BRs and two-body MMRs remains constant (see Section~\ref{subsec:scaling} for more details about the simulation setups). Despite the range of planet spacings and planet masses across the systems, the level of dynamical chaos experienced by nearly equally spaced systems (red boxes) is roughly constant: The fraction of initial conditions that produce MEGNO values ${>}\,3$ within each box ranges from $35$\,\% to $48$\,\%.}
\label{fig:chaos_grid}
\end{figure*}

If the transformed Hamiltonian is reduced to a single cosine term, either by discarding all other terms or utilizing a suitable approximation \citep[e.g.,][]{Hadden2019}, the system is integrable. However, if more than one cosine term is retained, the system may exhibit chaotic behavior, requiring more sophisticated analytical treatments or numerical methods. In this paper, we generate dynamical models that contain certain MMR (and/or secular) terms and numerically integrate them to compare directly with $N$-body simulations. For example, the Hamiltonian (or, more precisely, the energy) for a system of two planets that includes the Keplerian terms, first-order 7:6 MMR terms, and first-order 8:7 MMR terms would be written as
\begin{align} 
\label{Hexample}
H = -\frac{G M_* m_1}{2a_1} -\frac{G M_* m_2}{2a_2} &-\frac{G m_1 m_2}{a_{2,0}} \Big( \Big. 
\nonumber\\
\tilde{C}_{(7,-6,-1,0,0,0)}^{(0,0,0,0)}(\alpha_{1,2}) e_1 
&\cos(7\lambda_2 - 6\lambda_1 - \varpi_1) 
\nonumber\\ 
+\tilde{C}_{(7,-6,0,-1,0,0)}^{(0,0,0,0)}(\alpha_{1,2}) e_2 
&\cos(7\lambda_2 - 6\lambda_1 - \varpi_2) \nonumber\\
+ \tilde{C}_{(8,-7,-1,0,0,0)}^{(0,0,0,0)}(\alpha_{1,2}) e_1 &\cos(8\lambda_2 - 7\lambda_1 - \varpi_1)\nonumber\\ 
+ \tilde{C}_{(8,-7,0,-1,0,0)}^{(0,0,0,0)}(\alpha_{1,2}) e_2 
&\cos(8\lambda_2 - 7\lambda_1 - \varpi_2) \Big. \Big)
\end{align}
\noindent
where $a_{2,0}$ is a reference semi-major axis. Although concise in this form, note that $H$ must be expressed in terms of canonical coordinate-momentum pairs in order for Hamilton's equations to apply.

We integrate multiple ensembles of simplified dynamical models of five-planet systems that account for various combinations of resonant interactions. The equations of motion for each model are integrated with the eighth-order Dormand-Prince integrator \citep{Dormand&Prince1978}, \texttt{DOP853}, implemented in the \texttt{SciPy} package's \citep{Virtanen2020} \texttt{integrate.ode} class. We adopt a relative tolerance of $10^{-5}$, which achieves a similar energy error to full $N$-body simulations (see Appendix~\ref{sec:integrators} for more details). As in our $N$-body simulations, we monitor the planets' orbits in $10$,$000$ logarithmically spaced time intervals between $P_1$ and $10^7\,P_1$ and stop the simulation if the extent of two planets' orbits come within a specified distance threshold (the Hill radius of the innermost planet), the same as in our $N$-body simulations.

\section{$N$-body results}
\label{sec:results}

\subsection{Scaling with planet properties}
\label{subsec:scaling}

In Section~\ref{sec:discussion}, we will argue that instabilities in compact multiplanet systems are driven primarily by the overlap of a specific set of 3BRs. Here, we consider the question: If chaos is generated by the overlap of 3BRs, how does the degree of resonance overlap scale with planet mass, spacing, and eccentricity?

For a system of three planets with common period ratio, $\mathcal{P}$, and planet-to-star mass ratios, $m_i/M_*$, the sizes of the 3BRs, as measured by their fractional width in period ratio (or semi-major axis) space, scale linearly with the planet-to-star mass ratio. As we will demonstrate below, the dominant 3BRs are generated by combinations of the nearest two-body MMRs between the inner and outer planet pairs. Consider the set of 3BRs generated by combinations of two-body MMRs lying between adjacent $j$:$j-1$ and $j+1$:$j$ first-order MMRs, where $\frac{j+1}{j}\,{<}\,\mathcal{P}\,{<}\,\frac{j}{j-1}$. In other words, consider the 3BRs arising from combinations of two-body MMRs falling within one of the red boxes in Fig.~\ref{fig:chaos_grid}. The area of such a box, as measured in period ratio space, scales as $\paren{\frac{j}{j-1}\,{-}\,\frac{j+1}{j}}^2\approx{j^{-4}}$. Thus, the local density of 3BRs in a given box will scale as ${\propto}\,\fracbrac{m}{M_*}j^{4}$ or, in terms of the fractional separation in semi-major axes ($\Delta a/a\,{=}\,\frac{2}{3j}$), the 3BR density scales as ${\propto}\,\fracbrac{m}{M_*}\fracbrac{a}{\Delta a}^4$.

\citet{Hadden2018} showed that, for closely spaced planets with $\mathcal{P}\,{\lesssim}\,2$, the strengths of MMRs depend on orbital eccentricity only through the combination $e/e_\mathrm{cross}$. 3BRs are generated by the interactions of two-body MMRs between distinct planet pairs, so their strengths will therefore similarly depend on eccentricities only through the combination $e/e_\mathrm{cross}$.

According to the arguments given above, we expect the degree of resonance overlap and chaos to depend principally on only two parameters, $e/e_\mathrm{cross}$ and $\fracbrac{m}{M_*}\fracbrac{a}{\Delta a}^4$. To test this hypothesis, Fig.~\ref{fig:chaos_grid} shows a grid of $N$-body simulations of three-planet systems with different initial planet spacings, highlighting regions of regular and chaotic initial conditions. The eccentricities and planet masses have been scaled to maintain a constant 3BR density by keeping $e/e_\mathrm{cross}$ and $\fracbrac{m}{M_*}\fracbrac{a}{\Delta a}^4$ fixed. The systems are comprised of a test particle orbiting between an inner and outer planet. The orbital periods, $P_1$ and $P_3$, of the inner and outer planet are set as a multiple of the test-particle period, $P_2$, according to $P_1\,{=}\,\frac{J_1\,{-}\,1}{J_1}P_2$ and $P_3\,{=}\,\frac{J_2}{J_2\,{-}\,1}P_2$, where $J_1$ and $J_2$ were sampled uniformly on a grid of values in the interval $J_i\,{\in}\,[4,\,8]$. The planets' masses are chosen according to $m_i\,{=}\,5\,{\times}\,10^{-6} (4/J_i)^4 M_*$ for $i\,{=}\,1,\,3$ and $m_2\,{=}\,0$, where $M_*$ is the host-star mass. Orbital eccentricities are set so that $e_1\,{=}\,0.25(\frac{a_2}{a_1}\,{-}\,1)$, $e_2\,{=}\,0$, and $e_3\,{=}\,0.25(1\,{-}\,\frac{a_2}{a_3})$, where $a_i$ is the semi-major axis of the $i$th body. The initial angular orbital elements are set to $\lambda_1\,{=}\,\varpi_1\,{=}\,\pi/2$, $\lambda_2\,{=}\,0 $, and $\lambda_3\,{=}\,\varpi_3\,{=}\,3\pi/2$ where $\lambda_i$ and $\varpi_i$ are the mean longitude and longitude of periapsis, respectively, of the $i$th planet.

Figure~\ref{fig:chaos_grid} shows the results from an $800$\,$\times$\,$800$ grid of $N$-body simulations, colored to indicate chaotic (black) and regular (white) trajectories based on the value of the MEGNO chaos indicator, $Y$, computed from short integrations of length $5\,{\times}\,10^4\,P_2$. Specifically, the grayscale stretches from $Y_\mathrm{min}\,{=}\,2$ to $Y_\mathrm{max}\,{=}\,10$ so that all trajectories with Lyapunov times $t_\mathrm{Ly}/P_2\,{\lesssim}\,5\,{\times}\,10^4/Y_\mathrm{max}\,{=}\,5{\times}\,10^3$ appear as black. Despite the broad range of planet spacings and planet masses, Fig.~\ref{fig:chaos_grid} shows that these systems are subject to a consistent level of dynamical chaos, especially along the line $J_1=J_2$, which corresponds to equally spaced planets. Within the red boxes plotted in Fig.~\ref{fig:chaos_grid} to highlight (nearly) equally spaced systems, the fraction of initial conditions producing MEGNO values ${>}\,3$ ranges between $35$\,\% and $48$\,\%. Note that planets in typical exoplanet systems are approximately equally spaced; $84$\,\% of observed systems (taken from the NASA Exoplanet Archive\footnote{\url{exoplanetarchive.ipac.caltech.edu} (Accessed: 2024 May 8).}) containing three or more planets with compact spacings (i.e., $\mathcal{P}\,{<}\,2$ for each adjacent planet pair) fall in a red box from $J\,{=}\,2$ to $5$.

Below, we present the instability times for our nine $N$-body simulation ensembles, leveraging our theoretical understanding of how resonance overlap scales with the properties of the planets.

\begin{figure}
\centering
\includegraphics[width=0.40\textwidth]{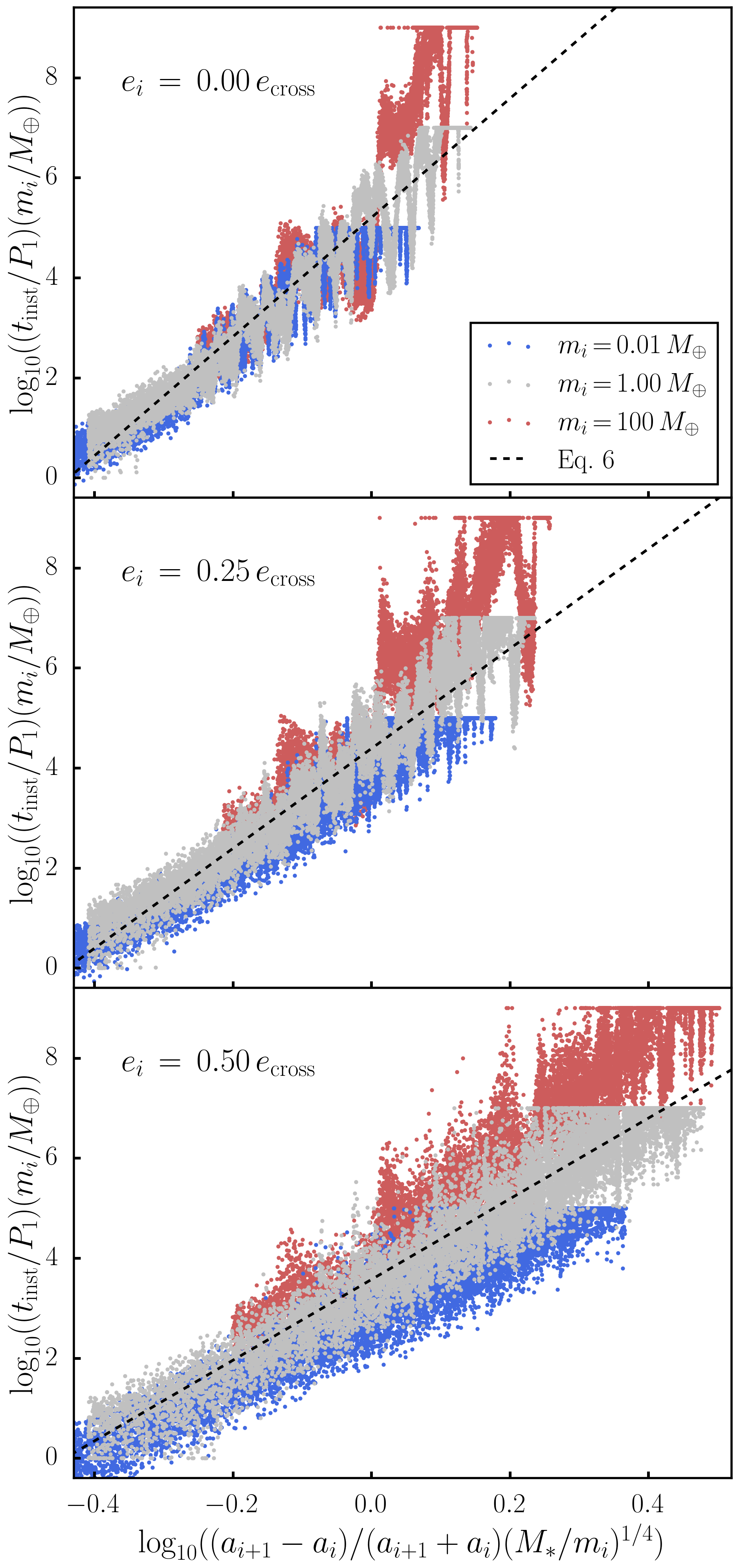}
\caption{Instability times from $N$-body simulations of initially equally spaced, five-planet systems. Systems with different planet masses are shown in different colors, and the three panels show systems initialized with different initial normalized eccentricities. Note the log-log axes, mass-dependent shift to the y-axis, and the factor of $\left({m_i}/{M_*}\right)^{-1/4}$ in the spacing units. The dotted black lines show the empirical fit given in Eq.~\ref{eq:linear_fit}.}
\label{fig:allsims}
\end{figure}

\subsection{Instability times}
\label{sec:nbodyresults}

Figure~\ref{fig:allsims} shows the results from our nine ensembles of $N$-body integrations. Each panel shows three ensembles at a common normalized eccentricity, spanning four orders of magnitude in planet mass, with instability time plotted against initial planet spacings scaled by a factor of $(m_i/M_*)^{1/4}$. This allows us to compare instability times across ensembles while holding the local densities of 3BRs fixed (see Section~\ref{subsec:scaling}). Instability times are scaled by $m_i/M_\oplus$, which comes from the analytic formula derived by \citet{Petit2020} for predicting the instability times of initially circular systems (their Eq.~83). We find, empirically, that this scaling holds for more eccentric systems as well. In these theoretically motivated units, instability times scale with initial planet spacing according to a power law, with parameters that depend only on the initial normalized eccentricity of the planets. We therefore perform an empirical fit to the instability times of the Earth-mass systems, adopting the functional form
\begin{multline}
\label{eq:linear_fit}
    \log_{10}\left(\frac{t_\mathrm{inst}}{P_1}\frac{m_i}{M_\oplus}\right) =\\  \paren{A + B\fracbrac{e_i}{e_\mathrm{cross}}}\log_{10}\left[{
        \frac{a_{i+1} - a_i}{a_{i+1} + a_i}
        \fracbrac{M_*}{m_i}^{1/4}
    }\right]\\
    + C + D \fracbrac{e_i}{e_\mathrm{cross}}~,
\end{multline}
where $A$, $B$, $C$, and $D$ are free parameters. Performing a least-squares fit,\footnote{We exclude values of $\frac{a_{i\,{+}\,1}\,{-}\,a_i}{a_{i\,{+}\,1}\,{+}\,a_i} \fracbrac{M_*}{m_i}^{1/4}$ for which ${>}\,10$\,percent of systems have $t_\mathrm{inst}\,{=}\,P_1$ or $t_\mathrm{inst}\,{=}\,10^7\,P_1$} we find $A\,{=}\,11.9$, $B\,{=}\,-7.67$, $C\,{=}\,5.20$, and $D\,{=}\,-3.26$. This empirical fit is plotted in each panel of Fig.~\ref{fig:allsims} and provides a good approximation to instability times over the plotted range of spacings, eccentricities, and masses.

\begin{figure*}
\centering
\includegraphics[width=\textwidth]{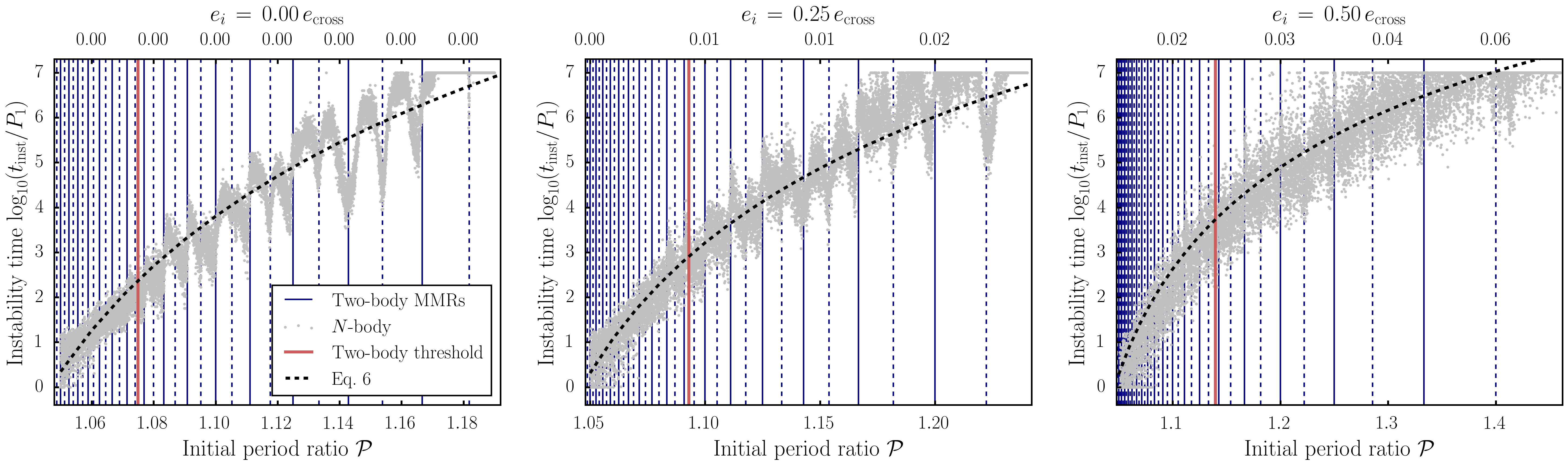}
\caption{Instability times from $N$-body simulations of systems comprised of five equally spaced, Earth-mass planets plotted versus initial adjacent-planet period ratio. Initial planet eccentricities, which were set to a fixed fraction of the orbit-crossing eccentricity in each panel, are indicated as a function of period ratio along the upper $x$-axis. The locations of first- and second-order MMRs are highlighted in blue, and the two-planet stability thresholds for initially circular \citep{Wisdom1980, Deck2013} and initially eccentric \citep{Hadden2018} systems are shown in red. The initially circular systems (left panel) are equivalent to the systems studied in \citet{Obertas2017}, in which the linear trend and dips at MMRs are discussed at length. As the normalized eccentricity increases, systems survive less long, with less significant dips in survival time at MMRs and larger overall spreads in survival time. Our power-law empirical fit (Eq.~\ref{eq:linear_fit}) predicts instability times well over a range of planet spacings and eccentricities.}
\label{fig:Nbodycompare}
\end{figure*}

Note that the above empirical fit assumes a power-law relationship between initial planet spacing and instability time, whereas previous works have typically adopted an exponential relationship \cite[e.g.,][]{Chambers1996, FaberQuillen2007, Smith&Lissauer2009, Obertas2017, Gratia&Lissauer2021}. For systems with low normalized eccentricities, both functional forms describe $N$-body results well. However, for systems with nonnegligible normalized eccentricities, a power-law relationship fits the $N$-body results significantly better than an exponential. This is visually apparent in Fig.~\ref{fig:Nbodycompare}: Over the range plotted, the period ratio is approximately given by $\mathcal{P}\,{\approx}\,1\,{+}\,3\frac{a_{i+1}\,{-}\,a_i}{a_{i+1}\,{+}\,a_i}$, so an exponential relationship between instability time and spacing would appear as a straight line in Fig.~\ref{fig:Nbodycompare}, which does not describe the $N$-body results well.

We can compare our empirical fit with that of previous works by considering the initially circular, Earth-mass planet case and linearizing the power-law fit. After linearizing about the mid-point of our range of spacing values, $\frac{a_{i+1}\,{-}\,a_i}{a_{i+1}\,{+}\,a_i} \fracbrac{M_*}{m_i}^{1/4}\,{=}\,0.74$, and converting to the functional form $\log(t_\mathrm{inst}/P_1)\,{=}\,b\Delta\,{+}\,c$, where $\Delta$ is the separation of the planets in mutual Hill radii, we calculate a slope $b\,{=}\,1.06$ and intercept $c\,{=}\,-1.52$. These values are comparable to the slopes and intercepts reported in previous works (e.g., $b\,{=}\,1.01$ and $c\,{=}\,-1.69$ from \citealt{Smith&Lissauer2009}).

Based on $N$-body simulations of nonequal-mass, nonequally spaced planets, \citet{Pu&Wu2015} reported that the instability times of systems comprised of slightly eccentric planets are approximately equal to those of initially circular systems if the planet separations are measured as the distance between the pericenter of the outer planet and the apocenter of the inner planet. Testing this claim with our $N$-body results, we find that applying our instability time fit (Eq.~\ref{eq:linear_fit}) with $e_i\,{=}\,0.0$ to eccentric systems with semi-major axis separations replaced by pericenter-apocenter separations systematically underpredicts the systems' instability times by a factor of ${\sim}\,10$\,--\,$100$. Our empirical fit predicts instability times much more accurately.

For the remainder of the paper, we will restrict ourselves to considering systems comprised of $m_i\,{=}\,1.00\,M_\oplus$ planets because our results can be easily generalized to other planet masses by rescaling the separations by a factor ${\propto}\,m_i^{1/4}$. Figure~\ref{fig:Nbodycompare} shows the instability times for our $N$-body ensembles of systems comprised of Earth-mass planets, now plotted against the planets' initial period ratios. Planets' normalized eccentricities increase across the panels from left to right, which lowers instability times. Period ratios where two-planet resonance overlap criteria would predict isolated pairs of planets initialized with the same masses and eccentricities to be chaotic are indicated by vertical lines in each panel. Clearly, these higher-multiplicity systems exhibit instabilities (and, as a corollary, dynamical chaos) at spacings well beyond these two-planet resonance overlap boundaries. Large dips in survival times are apparent in the initially circular simulations near the locations of first- and second-order MMRs. These dips were observed previously in simulations by \citet{Obertas2017}. Notice that dips in instability time near second-order MMRs occur precisely where $\mathcal{P}\,{=}\,p/(p\,-\,2)$ (dotted lines in Fig.~\ref{fig:Nbodycompare}), whereas dips near first-order MMRs occur at slightly smaller period ratios than $\mathcal{P}\,{=}\,p/(p\,-\,1)$ (solid lines in Fig.~\ref{fig:Nbodycompare}). These offsets were noticed but left unexplained in \citet{Obertas2017}. We show in Appendix~\ref{sec:MMRloc} that the offsets near first-order MMRs arise from the unique geometry of these resonances, which causes their separatrices to intersect $e\,{=}\,0$ slightly below $\mathcal{P}\,{=}\,p/(p\,-\,1)$. As planets' initial normalized eccentricities are increased, the dips in instability time near MMRs become less prominent.

\begin{figure*}
\centering
\includegraphics[width=\textwidth]{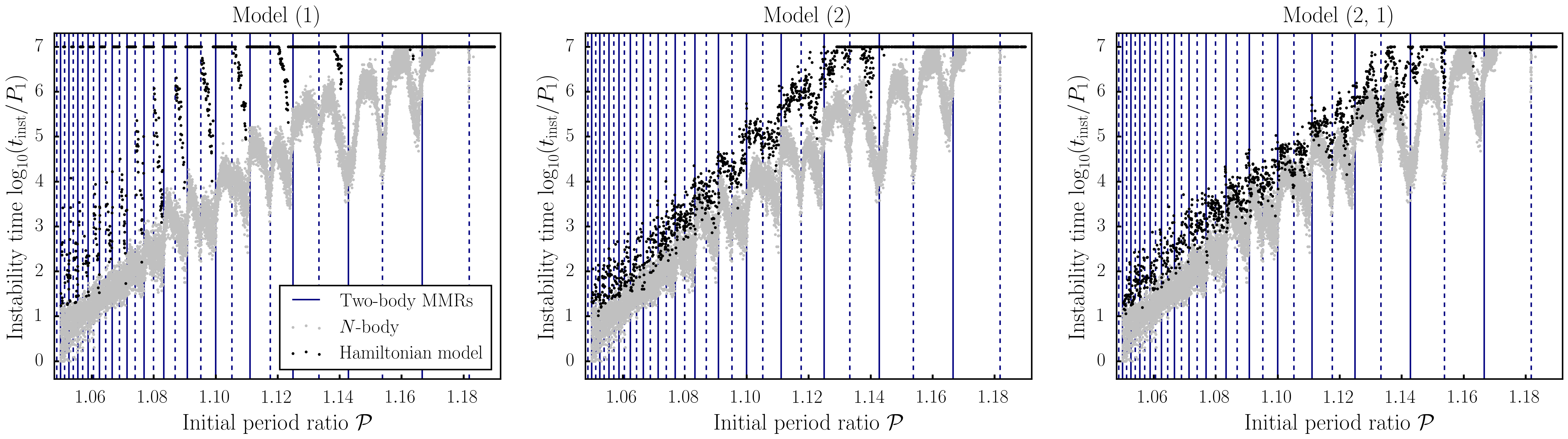}
\caption{Comparison of the instability times for $15$,$000$ initially circular, five-planet systems from $N$-body simulations (grey; leftmost plot in Fig.~\ref{fig:Nbodycompare}) and $1$,$500$ dynamical model simulations, which only include certain aspects of the dynamics (black). In the leftmost plot, we show the instability times predicted by a model that contains the nearest first-order MMR (solid blue lines) interactions between each adjacent planet pair (four total MMRs). The middle plot corresponds to a model that contains two first-order MMRs between adjacent planet pairs (eight total MMRs). The rightmost plot corresponds to a dynamical model that contains the two nearest first-order MMRs and the nearest second-order MMR (dotted blue lines) between adjacent planet pairs (12 total MMRs). Model (2) broadly reproduces the correct trend in instability times, and Model (2, 1) predicts instability times very accurately.}
\label{fig:e0.0MMRs}
\end{figure*}

\begin{figure}
\centering
\includegraphics[width=0.40\textwidth]{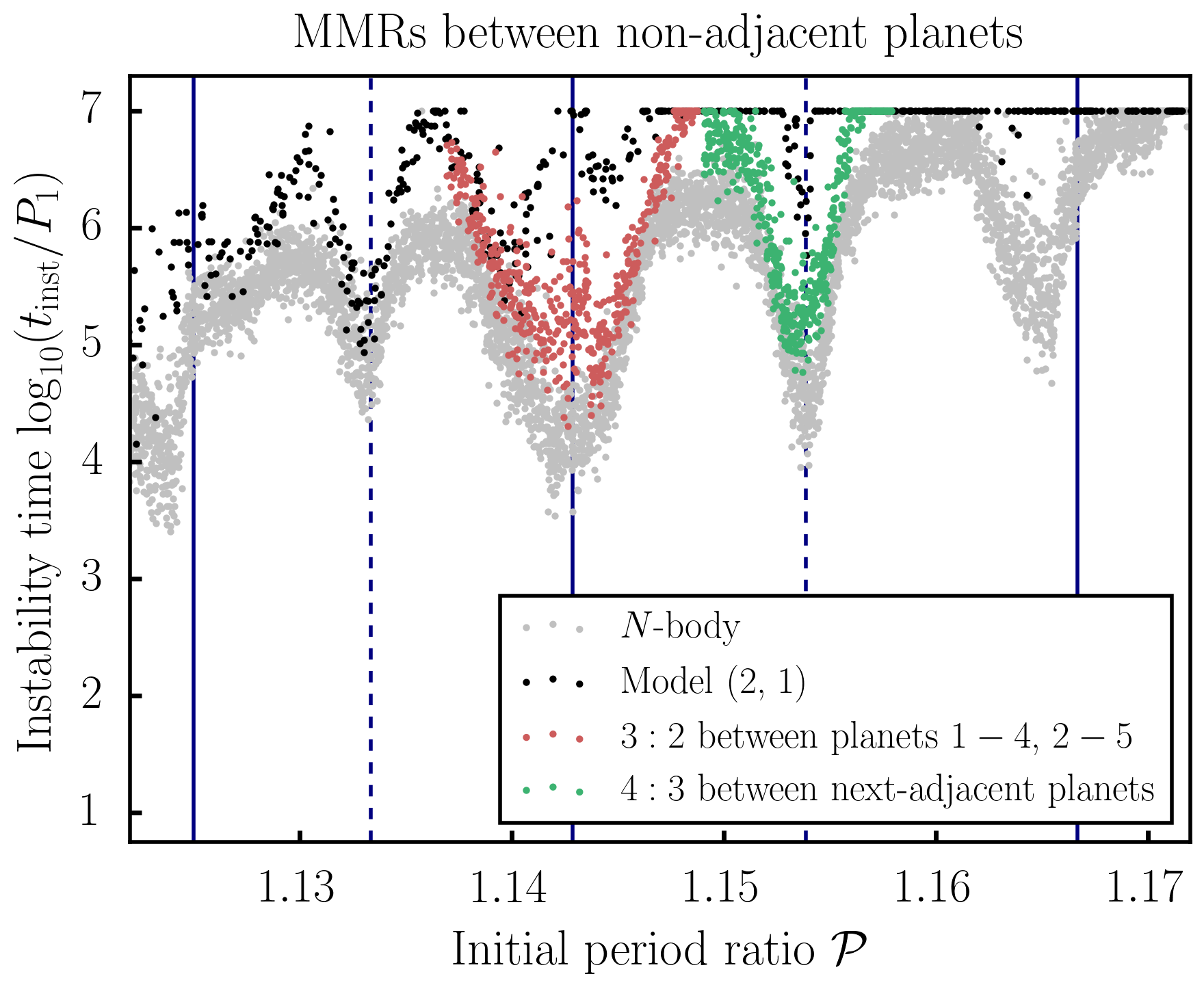}
\caption{Zoomed-in version of the right panel of Fig.~\ref{fig:e0.0MMRs}. Instability times predicted by a dynamical model that includes the 4:3 MMR between next-adjacent planets are shown in red, and instability times from a dynamical model that includes the 3:2 MMR between next-next-adjacent planets (planet pairs $1-2$ and $4-5$) are shown in green. These models precisely recover survival time around $\mathcal{P}\,{\approx}\,1.154$ and $\mathcal{P}\,{\approx}\,1.143$, respectively, which were missed by Model (2, 1). The improved agreement demonstrates that these dips in survival time are caused by first-order MMRs between nonadjacent planets rather than the MMRs between adjacent planets (the 15:13 and 8:7, respectively) that happen to lie nearby.}
\label{fig:e0.0otherMMRs}
\end{figure}

\section{Dynamical model results}
\label{sec:Hamiltonian_models}

We now turn to the results of our simplified dynamical model integrations. For each model considered, we generate ensembles comprised of $1$,$500$ equally spaced, five-planet systems following the same procedure we used to initialize $N$-body simulations described in Section~\ref{sec:simsetup}. We begin in Section~\ref{sec:circular} by comparing numerical models of increasing complexity against our $N$-body results for initially circular planetary systems. Then, in Section~\ref{sec:eccentric}, we turn to initially eccentric systems.

\subsection{Initially circular systems}
\label{sec:circular}

\begin{figure*}
\centering
\includegraphics[width=\textwidth]{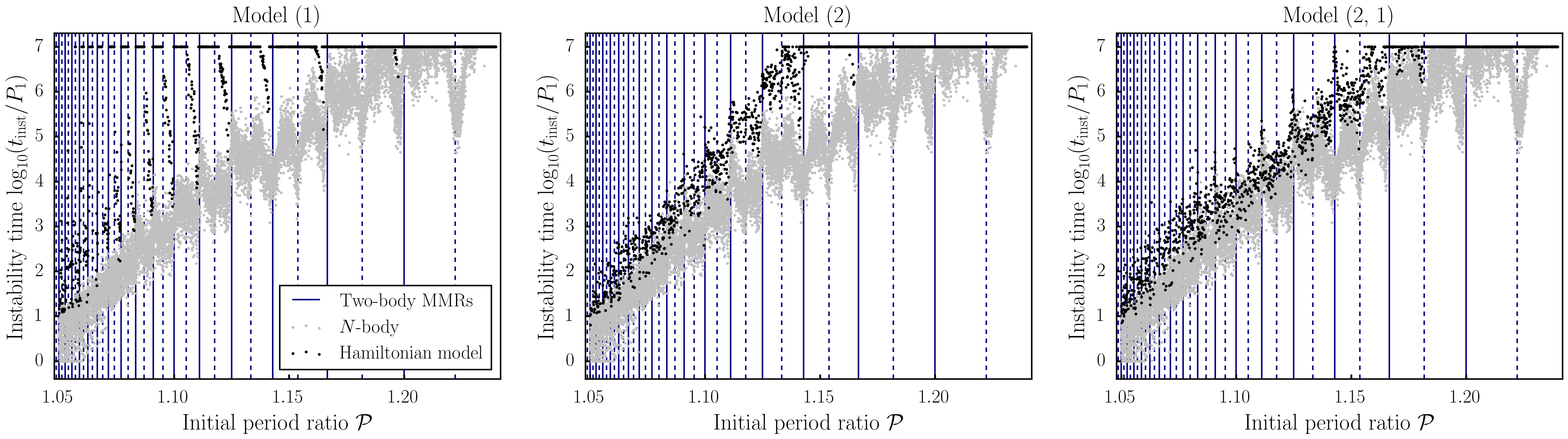}
\caption{Same comparison as Fig.~\ref{fig:e0.0MMRs}, now for the initially eccentric $e_i\,=\,0.25e_{\mathrm{cross}}$ systems (middle panel of Fig.~\ref{fig:Nbodycompare}). Similar to the initially circular case, Model (2, 1) predicts instability times fairly accurately when the planets begin with small normalized eccentricities.}
\label{fig:e0.25MMRs}
\end{figure*}

We begin by considering a series of three models of increasing complexity that account only for interactions between adjacent planets. The series of models are constructed as follows:
\begin{itemize}
\item 
    \textbf{Model (1)} includes a single first-order MMR interaction between each adjacent planet pair. For a system initialized with period ratio $\mathcal{P}$, we find the integer, $j$, satisfying 
    \begin{equation}
        j=\min_{k\in \mathbb{N}}\left|\mathcal{P} - \frac{k}{k-1}\right|
    \end{equation}
    and add the disturbing function terms associated with the resonant arguments $j\lambda_{i\,{+}\,1}\,{-}\,(j-1)\lambda_i\,{-}\,\varpi_i$ and $j\lambda_{i\,{+}\,1}\,{-}\,(j-1)\lambda_i\,{-}\,\varpi_{i\,{+}\,1}$ with $i=1,\,2,\,3,\,\mathrm{and}~4$. Consequently, this model contains a total of $(4\,\mathrm{pairs})\,\times\,(2\,\mathrm{cosine~arguments})\,{=}\,8$ cosine terms.
\item 
    \textbf{Model (2)} includes two first-order MMR interactions for each pair of adjacent planets. Specifically, given an initial period ratio $\mathcal{P}$ we find the integer $j$ such that $\frac{j+1}{j}\,{<}\,\mathcal{P}\,{<}\,\frac{j}{j-1}$ and add all disturbing function terms at first order in eccentricity associated with both the $j$:$j-1$ and $j+1$:$j$ MMRs to our Hamiltonian. This model contains a total of $16$ cosine terms.
\item 
    \textbf{Model (2, 1)} extends Model (2) by additionally including disturbing function terms associated with the nearest second-order MMR for each adjacent planet pair. In other words, we find the integer $j$ for which
    \begin{equation}
        j=\min_{k\in \mathbb{N}}\left|\mathcal{P} - \frac{k}{k-2}\right|
    \end{equation}
    and add the disturbing function terms associated with the resonant arguments $\theta_\mathrm{res}\,{-}\,2\varpi_i,\theta_\mathrm{res}\,{-}\,\varpi_i\,{-}\,\varpi_{i\,{+}\,1},~\mathrm{and}~\theta_\mathrm{res}\,{-}\,2\varpi_{i\,{+}\,1}$, where $\theta_\mathrm{res}\,{=}\,j\lambda_{i\,{+}\,1}\,{-}\,(j\,{-}\,2)\lambda_i$. The resulting model contains a total of $16\,{+}\,4\,{\times}\,3\,{=}\,28$ cosine terms.
\end{itemize}

We emphasize that the resonances included in each of our dynamical models are selected based on the {\it initial} system configurations, and are not updated as the spacing of the planets change.

Figure~\ref{fig:e0.0MMRs} compares the instability time scales measured for this series of models against the $N$-body integrations presented in Section~\ref{sec:results}. Model (1), shown in the leftmost panel of Fig.~\ref{fig:e0.0MMRs}, only predicts instabilities to occur when systems are initialized in the close vicinity of a first-order MMR. While the $N$-body integrations similarly show dips in instability time in the vicinity of first-order MMRs, Model (1) is clearly inadequate for explaining the overall trend in instability times.

Instability times for Model (2) are shown in the middle panel of Fig.~\ref{fig:e0.0MMRs}. The additional first-order MMR interactions dramatically improve the agreement of this model with $N$-body results. Both the overall trend, as well as many of the dips at first-order MMRs observed in the $N$-body integrations, are reproduced by this model.

Instability times for Model (2, 1) are shown in the rightmost panel of Fig.~\ref{fig:e0.0MMRs}. The addition of second-order MMRs further improves the agreement between model instability times and $N$-body results, especially at wide separations.

While Model (2) and Model (2, 1) both broadly reproduce $N$-body instability times, there are multiple significant dips of ${\sim}\,2$ orders of magnitude at period ratios $\mathcal{P}\,{>}\,1.14$ that they fail to reproduce. We show, with integrations of additional models, that these dips can be attributed to resonant interactions between nonadjacent planet pairs in the system.

Figure~\ref{fig:e0.0otherMMRs} shows the instability time predictions, in restricted ranges of initial planet spacing, for two models that extend Model (2, 1) by accounting for first-order MMR interactions between nonadjacent planets. The first model, shown in green in Fig.~\ref{fig:e0.0otherMMRs}, includes the 4:3 MMR interactions between each next-adjacent planet pair (i.e., the first and third, second and fourth, and third and fifth planets). This model successfully reproduces the large drop in instability times centered near $\mathcal{P}\,{=}\,\fracbrac{4}{3}^{1/2}\,{\approx}\,1.155$. We also consider a model, shown in red in Fig.~\ref{fig:e0.0otherMMRs}, that extends Model (2, 1) by additionally including the 3:2 MMR interactions between the first and fourth as well as the second and fifth planets. Integrations with this model reproduce the significant dip in instability time observed near $\mathcal{P}\,{=}\,\fracbrac{3}{2}^{1/3}\,{\approx}\,1.144$.

\subsection{Eccentric systems}
\label{sec:eccentric}

We now turn to initially eccentric systems. Figure~\ref{fig:e0.25MMRs} shows a series of comparisons between $N$-body instability times and Models (1), (2), and (2, 1) for systems with initial planet eccentricities set to $0.25e_\mathrm{cross}$. As in the initially circular case shown in Fig.~\ref{fig:e0.0MMRs}, including a single MMR between adjacent planet pairs only results in instabilities in the close vicinity of two-body MMRs. Again, both Model (2) and Model (2, 1) predict that instability times increase with the initial planet spacings in a way that is broadly consistent with the $N$-body results. However, compared to the initially circular systems shown in Fig.~\ref{fig:e0.0MMRs}, a more significant improvement in agreement with $N$-body results is obtained going from Model (2) to Model (2, 1). This improvement underscores the increasing dynamical importance of higher-order MMRs at larger eccentricity values.

\begin{figure*}
\centering
\includegraphics[width=\textwidth]{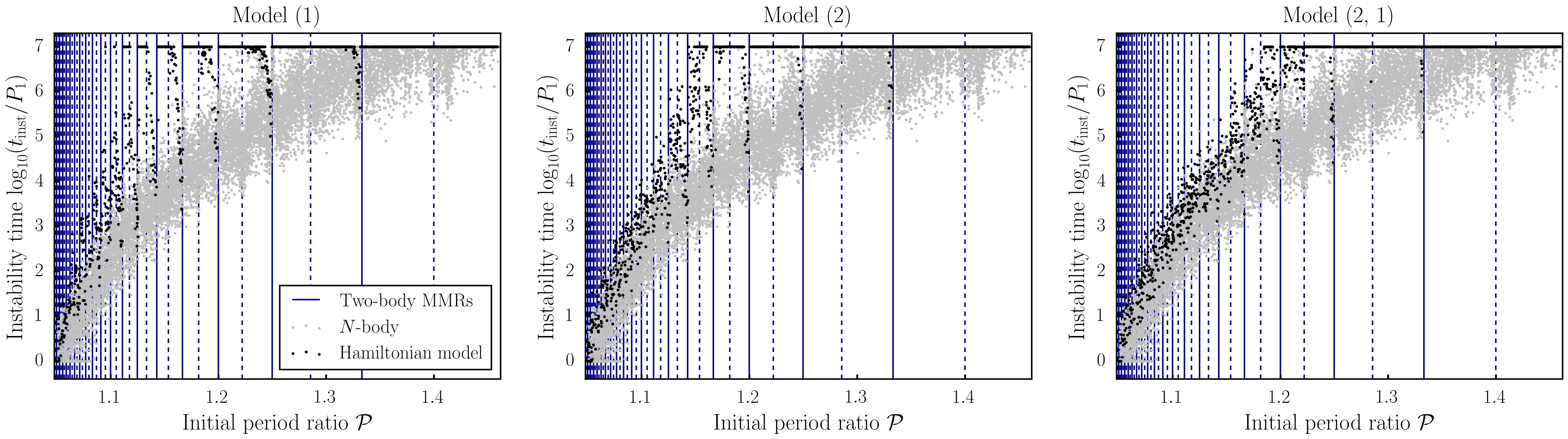}
\caption{Same comparison as Fig.~\ref{fig:e0.0MMRs}, now for the initially eccentric $e_i\,=\,0.50e_{\mathrm{cross}}$ systems (right panel of Fig.~\ref{fig:Nbodycompare}). For systems with larger normalized eccentricities, Model (2, 1) is no longer sufficient to recover the true instability time trends, especially at large initial planet spacings.}
\label{fig:e0.5MMRs}
\end{figure*}

\begin{figure*}
\centering
\includegraphics[width=0.75\textwidth]{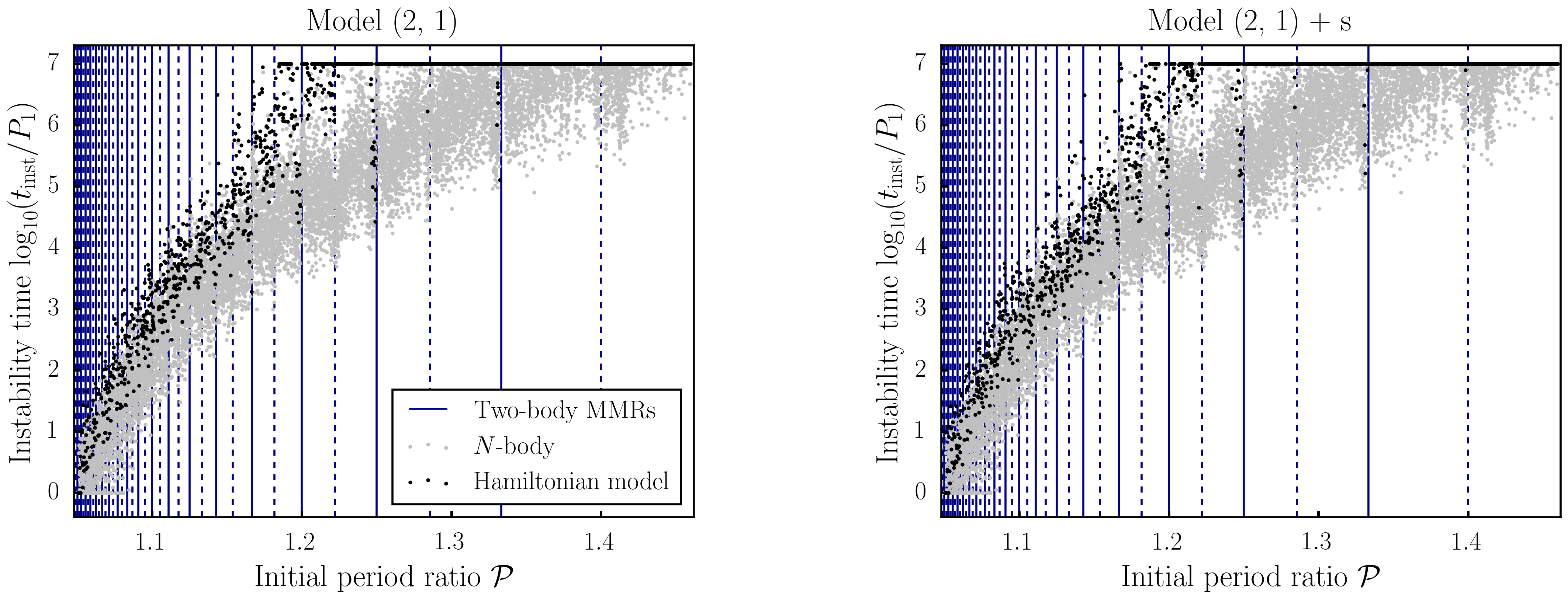}
\caption{Comparison between the instability times predicted by Model (2, 1) (as in the right panel Fig.~\ref{fig:e0.5MMRs}) and the instability times predicted by Model (2, 1) + s, which additionally includes second order secular terms between all planet pairs. Including secular evolution in the model does not meaningfully affect the predicted instability times (similarly, including secular evolution does not affect the predictions of Model (2, 1, 1, 1) in Fig.~\ref{fig:e0.5moreMMRs}).}
\label{fig:e0.5sec}
\end{figure*}

\begin{figure*}
\centering
\includegraphics[width=\textwidth]{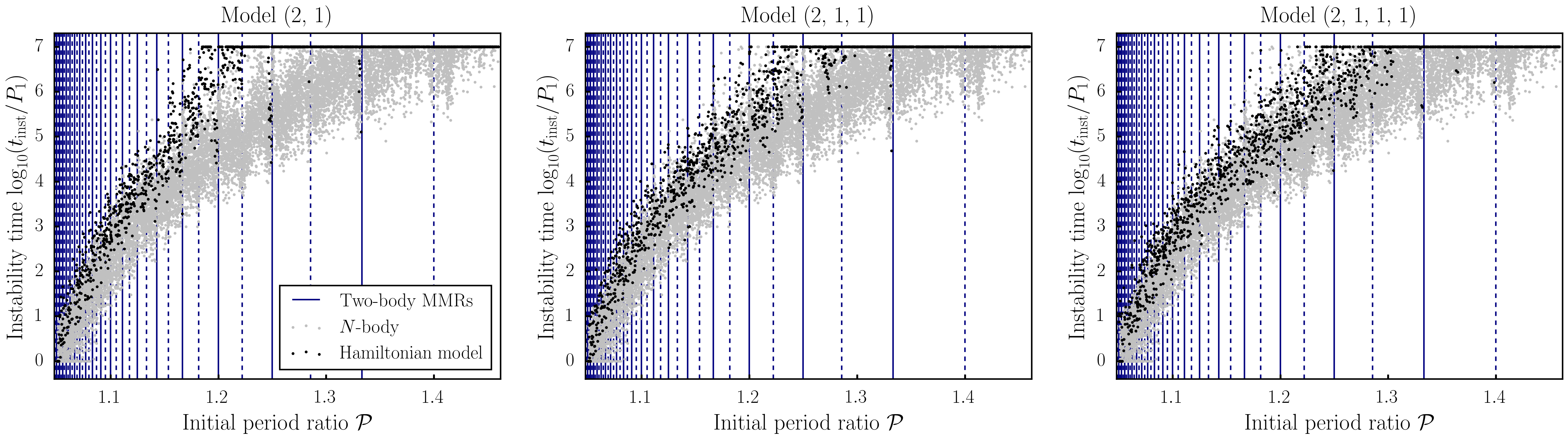}
\caption{Same comparison as Fig.~\ref{fig:e0.5MMRs}, now for models that include more MMRs between adjacent planet pairs. Predictions from Model (2, 1) are shown in the left panel, predictions from Model (2, 1, 1) are shown in the middle panel, and the right panel shows predictions from Model (2, 1, 1, 1). Including additional higher-order MMRs between adjacent planets improves the agreement with $N$-body instability times, particularly for systems initialized with wider planet spacings.}
\label{fig:e0.5moreMMRs}
\end{figure*}

Figure~\ref{fig:e0.5MMRs} shows instability times predicted by Models (1), (2), and (2, 1) for initial eccentricities of $0.50e_\mathrm{cross}$. Although including additional MMRs between adjacent planet pairs improves the agreement with $N$-body results, Model (2, 1) incorrectly predicts that systems with initial spacing $\mathcal{P}\,{\gtrsim}\,1.2$ should survive longer than $10^7\,P_1$ integrations. This suggests that higher-order resonances and/or secular evolution play a significant role in the dynamics of these systems.

To explore the influence of both secular terms and higher-order MMRs, we construct three additional dynamical models:
\begin{itemize}
\item 
    \textbf{Model (2, 1) + s} extends Model (2, 1) by adding the leading-order secular terms between all planet pairs. These secular terms are comprised of three individual terms, ${\propto}\,e_i^2$, ${\propto}\,e_{i\,{+}\,1}^2$, and ${\propto}\,e_i e_{i\,{+}\,1}\cos(\varpi_i\,{-}\,\varpi_{i\,{+}\,1})$ for each planet pair.
\item 
    \textbf{Model (2, 1, 1)} extends Model (2, 1) by including the nearest third-order MMR between each adjacent planet.
\item 
\textbf{Model (2, 1, 1, 1)} extends model (2, 1, 1) by adding the nearest fourth-order MMR terms between each adjacent planet pair.
\end{itemize}

Figure~\ref{fig:e0.5sec} compares Models (2, 1) and (2, 1) + s, clearly demonstrating that the inclusion of secular terms has no influence on the predicted instability times. We have also conducted tests with dynamical models containing secular terms up to fourth order in planet eccentricities, which introduces the possibility of nonlinear secular resonances and chaos. We find, however, that the inclusion of these higher-order secular terms also has a negligible influence on instability times.

Figure~\ref{fig:e0.5moreMMRs} illustrates the influence of higher-order MMRs, comparing instability time scale results for Models (2, 1), (2, 1, 1), and (2, 1, 1, 1). Clearly, including higher-order MMRs progressively improves the agreement with the $N$-body results, especially for systems with wide initial separations.

We have performed some numerical experiments with systems of nonequally spaced planets, and we find that the agreement between the dynamical model and $N$-body instability times is typically similar to the equally spaced case.

\section{Discussion}
\label{sec:discussion}

\subsection{Application to observed multiplanet systems}

The numerical results presented above have implications for understanding the processes that shape the observed population of compact multiplanet systems. In particular, the spacing of planets in the synthetic systems presented in Fig.~\ref{fig:allsims} and Fig.~\ref{fig:Nbodycompare} extend to that of tightly spaced observed systems. For eccentricities typical in these systems \citep[$e\,{\sim}\,0.02$; e.g.,][]{Hadden&Lithwick2017}, our empirical fit (Eq.~\ref{eq:linear_fit}) indicates that planets with mass $m\,{\sim}\,M_\oplus$ must be spaced with $\mathcal{P}\,{\gtrsim}\,1.35$ to be stable over $10^9\,P_1$.\footnote{We have confirmed that our empirical fit holds up to $10^9\,P_1$ for the $e_i\,{=}\,0.25\,e_{\mathrm{cross}}$ and $e_i\,{=}\,0.50\,e_{\mathrm{cross}}$ systems. Widely spaced, initially circular systems eventually experience a sharp increase in instability time (at $\mathcal{P}\,\approx\,1.17$; see \citealt{Obertas2017}), beyond which the empirical fit no longer holds. The empirical fit also breaks down for very eccentric systems ($e_i\,{\gtrsim}\,0.7\,e_{\mathrm{cross}}$), at which point secular evolution causes inevitable orbit crossing.} This agrees with the orbital spacing below which the prevalence of observed systems containing three or more planets falls off \citep{Weiss2018}, supporting the hypothesis that observed systems have been shaped by previous dynamical instabilities.

Our numerical results can also be leveraged to place eccentricity constraints on observed systems by requiring long-term stability. Based on our empirical fit, systems with compact planet spacings ($\mathcal{P}\,{\lesssim}\,1.5$) must possess very low eccentricities ($e\,{\lesssim}\,0.05$) to be stable for $10^9\,P_1$. Note that the constraints stated above come from our empirical fit to $N$-body simulations of five-planet systems, but instability times for $N\,{>}\,2$ systems are known to depend only weakly on $N$ \citep{Chambers1996, Petit2020}. Stability constraints for systems with different planet masses, spacings, and eccentricities can be deduced from our empirical fit by requiring that $t_\mathrm{inst}\,{\lesssim}\,10^9\,P_1$.

The fact that low-order MMRs between nonadjacent planets can significantly influence stability (Fig.~\ref{fig:e0.0otherMMRs}) suggests that observed multiplanet systems may also be sculpted by these resonances. For instance, the 2:1 MMR between next-adjacent planet pairs occurs at planet spacings that are common in compact observed systems ($\mathcal{P}\,{\approx}\,1.414$ if the planets are equally spaced). Additionally, we caution that due to the influence of MMRs between distant planet pairs, it is not always reasonable to separate a system consisting of many planets into adjacent sub-trios (e.g., for training machine learning models; \citealt{Tamayo2020}).

\subsection{Three-body resonances and the instability mechanism}
\label{sec:3BRs}

Our dynamical model simulations show a stark transition in instability behavior between Model (1), which includes only one set of first-order MMR terms per adjacent planet pair, and Model (2), which includes two sets of MMR terms for each adjacent planet pair. We argue below that this transition in behavior reflects the result of overlap between 3BRs in Model (2) that cannot arise in Model (1).

In Appendix~\ref{appendix:3br_widths}, we show that resonance terms of order $e^{k'}$ in eccentricity involving the mean longitude combination $j'\lambda_i\,{+}\,(k'\,{-}\,j')\lambda_{i\,{-}\,1}$ for an inner pair of planets and terms of order $e^{k}$ in eccentricity involving the combination $j\lambda_{i\,{+}\,1}\,{+}\,(k\,{-}\,j)\lambda_{i}$ for an outer pair of planets produce three-body terms involving the combination $j\lambda_{i\,{+}\,1}\,{+}\,(k-j\,{-}\,j')\lambda_{i}\,{+}\,(j'\,{-}\,k')\lambda_{i\,{-}\,1}$ that are of order $e^{k\,{+}\,k'\,{-}\,2}$ in eccentricities as well as three-body terms involving the mean longitude combination $j\lambda_{i\,{+}\,1}\,{+}\,(k\,{-}\,j\,{+}\,j')\lambda_{i}\,{+}\,(k'\,{-}\,j')\lambda_{i\,{-}\,1}$ that are of order $e^{k\,{+}\,k'}$. Details on computing the widths of these 3BRs are given in Appendix~\ref{appendix:3br_widths}.

Figure~\ref{fig:mmr_web} shows the extents of all 3BRs and two-body MMRs occurring between adjacent planets in Model (2, 1) for systems with eccentricities set to $e\,{=}\,0.25e_\mathrm{cross}$ and period ratios between the 8:7 and 9:8 MMRs. The extents of two-body MMRs in Fig.~\ref{fig:mmr_web} are indicated by vertical and horizontal gray bands, while the extents of various classes of 3BRs are indicated by different colored regions. The widths of these MMRs depend on planets' relative orbital orientations, so the dark and light shading in Fig.~\ref{fig:mmr_web} indicates median and maximum MMR extents, respectively, computed from a sample of $500$ random initial orientations. The highlighted widths provide an estimate of the initial extents of resonant regions, but note that as planets' eccentricities evolve, the extent of regions associated with resonances can vary.

The largest 3BRs appearing in Fig.~\ref{fig:mmr_web} (shown in red) are the ``zeroth-order" resonances involving resonant angles $j\lambda_{i\,{+}\,1}\,{+}\,(1\,{-}\,2j)\lambda_{i}\,{+}\,(j\,{-}\,1)\lambda_{i\,{-}\,1}$ generated by pairs of first-order MMRs. For Model (1), only one such 3BR will be present for each adjacent trio of planets, while in Model (2), there will be two such 3BRs, introducing the potential for these 3BRs to overlap and drive chaos; we attribute the qualitative change in instability behavior between Model (1) and Model (2) to the potential for these two 3BRs to overlap. Model (2) possesses the additional 3BRs indicated in purple in Fig.~\ref{fig:mmr_web}. These are the 3BRs arising as combinations of two first-order MMR angles in the form $(j\lambda_3\,{-}\,(j\,{-}\,1)\lambda_2)\,{+}\,(j'\lambda_2\,{-}\,(j'\,{-}\,1)\lambda_1)$. The strengths of these 3BRs scale with the planets' eccentricities as $e^2$, making them weaker than the zeroth-order 3BRs.

The introduction of second-order MMR terms in Model (2, 1) produces additional 3BRs with nonnegligible widths. The extents of these 3BRs are shown in blue, green, and orange in Fig.~\ref{fig:mmr_web}. Blue regions indicate 3BRs generated by combinations of a first-order two-body MMR of one planet pair and a second-order two-body MMR of the other. The green regions show the extent of the 3BR generated by the combination of the two second-order 17:15 MMRs between the inner and outer planet pair. Finally, the orange regions show the 3BRs generated as combinations of first-order and second-order MMR angles of the form $(j\lambda_3\,{-}\,(j\,{-}\,2)\lambda_2)\,{+}\,(j'\lambda_2\,{-}\,(j'\,{-}\,1)\lambda_1)$. The strengths of these 3BRs scale with the planets' eccentricities as $e^3$.

\begin{figure}
    \centering
    \includegraphics[width=0.47\textwidth]{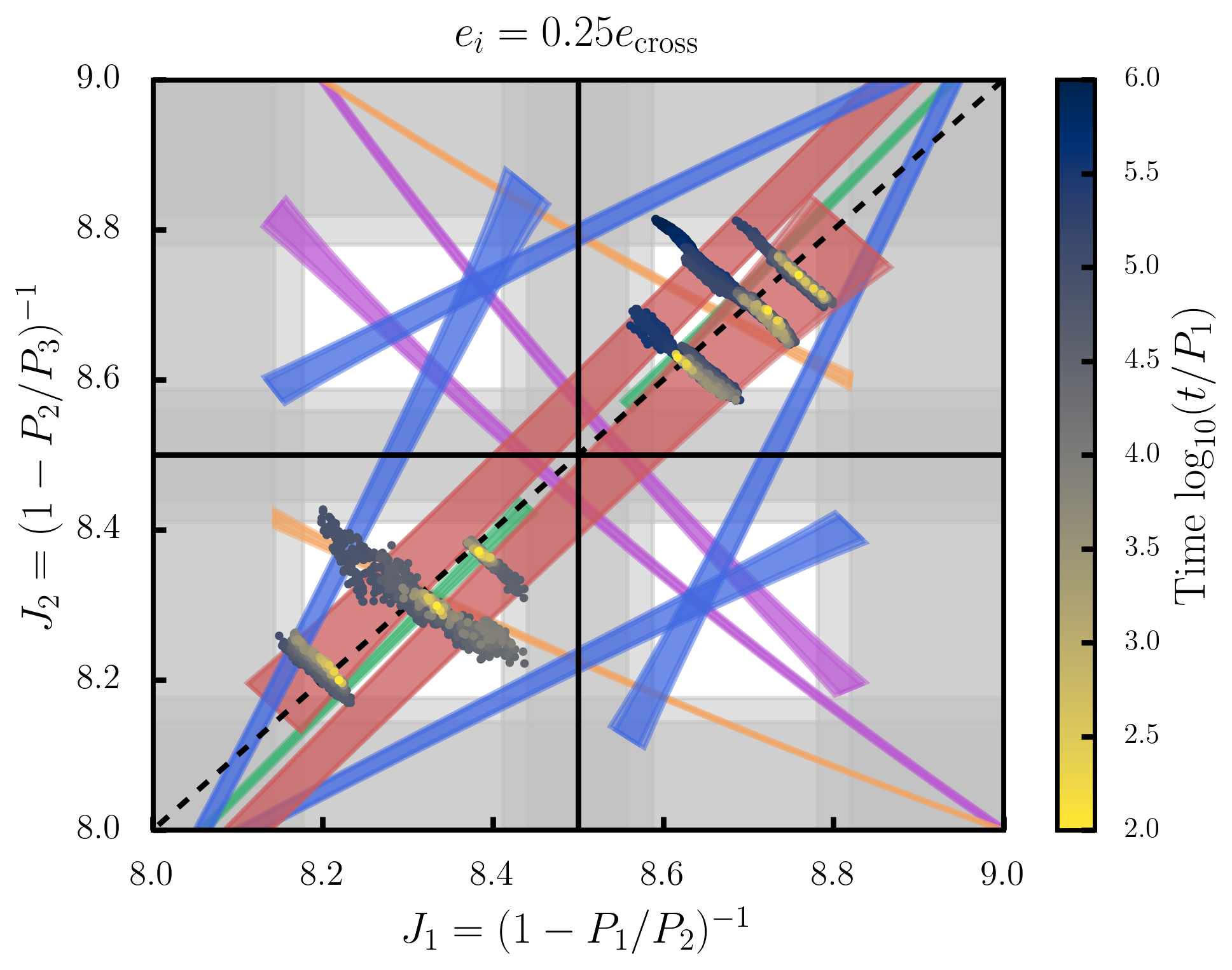}
    \caption{The web of 3BRs and two-body MMRs for three-planet systems located between the 8:7 and 9:8 MMRs with eccentricities equal to $0.25e_\mathrm{cross}$. Example trajectories through period ratio space, as predicted by Model (2, 1), are plotted on top of the resonance web (trajectories are truncated once they reach a two-body MMR, after which they begin rapidly wandering through period ratio space; see Section~\ref{sec:3BRs}). Overlap of the 3BRs shown in red and green causes the planets' period ratios to diffuse slowly, which eventually places the system in a two-body MMR, after which the system is short lived.}
    \label{fig:mmr_web}
\end{figure}

In addition to the MMR and 3BR widths, we plot the results of Model (2, 1) integrations of three-planet systems initialized with equally spaced period ratios, randomized orbital phases and orientations, and eccentricities set to $0.25e_\mathrm{cross}$. Each initial condition is integrated until the system enters the 8:7, 9:8, or 17:15 two-body MMRs, after which they begin rapidly wandering period ratio space, resulting in a close encounter. To place the example trajectories on Fig.~\ref{fig:mmr_web}, we use ``mean," rather than osculating, orbital parameters, derived via a near-identity canonical transformation using \texttt{celmech}'s \texttt{FirstOrderGeneratingFunction} class \citep{celmech2022}. We choose our generating function to eliminate the 8:7, 9:8, and 17:15 two-body MMR terms included in Model (2, 1) to first order in the planet masses. In this way, Fig.~\ref{fig:mmr_web} illustrates how the overlap of specific 3BRs dictates the chaotic transport of these systems into major two-body MMRs.

\begin{figure*}
\centering
\includegraphics[width=0.93\textwidth]{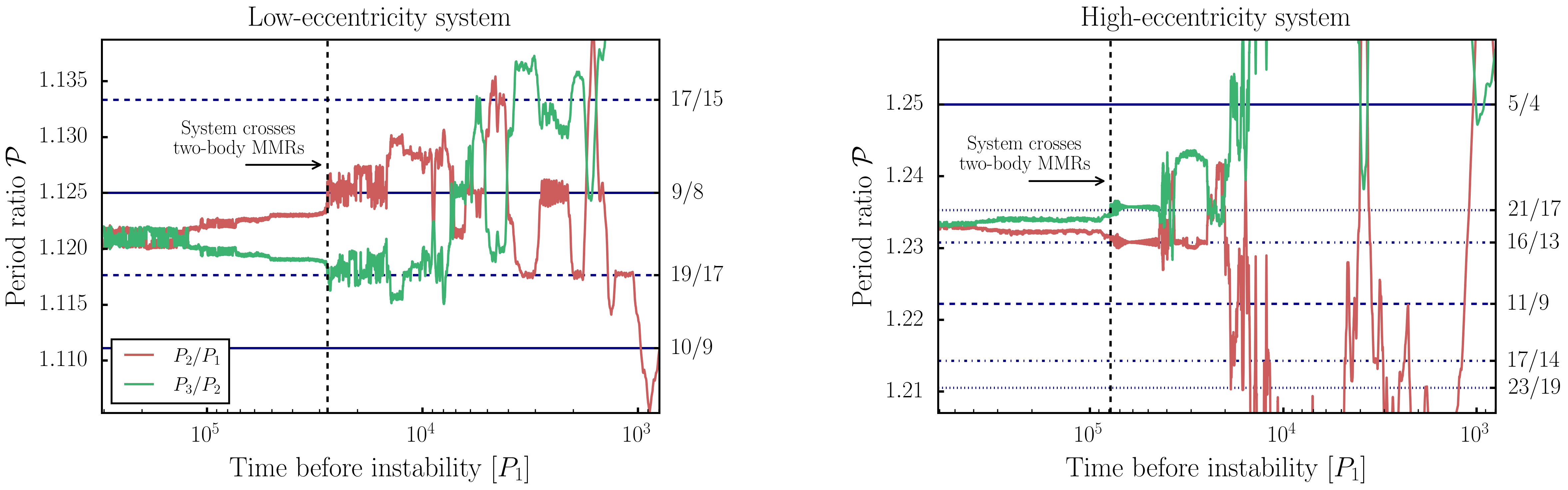}
\caption{Evolution of the planet period ratios during an $N$-body simulation of an initially circular three-planet system (left panel) and a high-eccentricity three-planet system ($e_i\,{=}\,0.50e_\mathrm{cross}$; right panel). To illustrate the chaotic diffusion that dominates the systems' lifetimes, we apply a boxcar average over a window of $100\,P_1$ and adopt a logarithmic x-axis. For low-eccentricity systems, the final instability phase is typically initiated by crossing a first-/second-order MMR (see also Fig.~2 in \citealt{Petit2020}). However, for high-eccentricity systems, higher-order MMRs can also initiate instability. This explains, in part, why including higher-order MMRs improves instability time predictions for high-eccentricity systems (Fig.~\ref{fig:e0.5moreMMRs}).}
\label{fig:example}
\end{figure*}

It is also apparent from Fig.~\ref{fig:mmr_web} that the slow chaotic drift caused by the overlap of 3BRs dominates the lifetimes of eventually unstable systems. To further illustrate this, Fig.~\ref{fig:example} shows $N$-body integrations of two unstable three-planet systems, one with initially circular orbits and one with eccentric orbits. Each panel shows the evolution of the planet period ratios, with a boxcar average applied over a window of $100\,P_1$. The period ratios in both panels exhibit a prolonged period of slow drift until one of the planet pairs encounters a strong two-body MMR.

The initially circular system shown in the left panel of Fig.~\ref{fig:example} agrees with the description of chaotic transport for initially circular, compact, three-planet systems given by \citet{Petit2020}. In their picture, overlapping zeroth-order 3BRs cause diffusion of the planet period ratios in the direction transverse to the (nearly) parallel zeroth-order 3BRs. This 3BR-driven diffusion eventually causes an adjacent planet pair to encounter a first-order two-body MMR, which destabilizes the system. Although \citet{Petit2020} consider the chaotic diffusion driven by the complete network comprised of all zeroth-order 3BRs generated by pairs of two-body MMRs, the agreement of our Model (2) results with $N$-body simulations indicates that the chaotic diffusion is dominated by the two zeroth-order 3BRs generated by the nearest pairs of first-order MMRs (shown in red in Fig.~\ref{fig:mmr_web}).\footnote{\citet{Petit2020} also consider contributions to zeroth-order 3BR amplitudes that arise at second order in planet masses from combinations of two-body disturbing function terms that are zeroth order in eccentricity. We find that the relative contributions of these terms to the full zeroth-order 3BR strengths are nearly always negligible (see Appendix~\ref{appendix:3br_widths})}.

The right panel of Fig.~\ref{fig:example} shows the evolution of an initially eccentric three-planet system. The introduction of nonzero eccentricities adds multiple complications to the picture of 3BR-driven chaotic diffusion described above. First, the strengths of zeroth-order 3BRs can be enhanced and evolve with planet eccentricities (see Eq.~\ref{eq:3br_amp_m}). Second, many additional higher-order 3BRs will have nonnegligible widths, as shown in Fig.~\ref{fig:mmr_web}. Third, surrounding higher-order MMRs will also have nonnegligible widths. If these higher-order MMRs are sufficiently strong to destabilize the system when a planet pair encounters them, then the distance that systems experiencing 3BR-driven chaotic diffusion must travel to destabilize can be substantially reduced. This effect is apparent in the $N$-body integration shown in the right panel of Fig.~\ref{fig:example}, which destabilizes shortly after the inner and outer planet pairs encounter the 16:13 and 21:17 MMRs, respectively. The relevance of higher-order MMRs explains why, in contrast with low-eccentricity systems, the instability times of eccentric systems do not have dips at the location of low-order MMRs (see Fig.~\ref{fig:Nbodycompare}). We note that it is not immediately clear what determines whether a two-body MMR is ``sufficiently strong" to precipitate the final instability phase. We defer detailed analysis of this question and other aspects of chaotic transport in eccentric systems to future work.

\section{Summary and conclusions}
\label{sec:conclusion}

Despite much previous work on the stability of multiplanet systems, the dynamical mechanism by which compact (exo)planetary systems destabilize remains somewhat unclear. Theoretical progress has been hindered by the difficulty of predicting instability times, which quickly becomes analytically intractable when considering the influence of many resonances. In this work, we pursue a semi-analytic approach, enabled by the open-source \celmech code \citep{celmech2022}, in which we generate dynamical models that account for various combinations of resonant interactions between the planets. By comparing numerical integrations of the simplified dynamical models with $N$-body simulations, we are able to clarify the mechanism by which compact planetary systems destabilize.

Our simplified dynamical models reveal that, surprisingly, $N$-body instability times can be accurately predicted by considering the influence of only a handful of relevant nearby MMRs. This points toward a remarkably simple physical picture in which specific, identifiable resonances overlap and drive chaos. We argue that the dominant resonances are 3BRs, generated by nearby two-body MMRs, which overlap in compact systems, causing a slow chaotic diffusion in the period ratios of the planets. Once a pair of planets enters a sufficiently strong two-body MMR, the system destabilizes. For low-eccentricity systems, first-/second-order MMRs are typically required to initiate the final instability phase, whereas higher-order MMRs can destabilize high-eccentricity systems (in some cases, low-order MMRs between nonadjacent planet pairs can also initiate instability).

Leveraging our understanding of the instability mechanism, we point out that the density of relevant 3BRs scales with planet mass and spacing as ${\propto}\,\fracbrac{m}{M_*}\fracbrac{a}{\Delta a}^4$ and depends on eccentricity only through the combination $e/e_\mathrm{cross}$. Based on nine ensembles of $N$-body simulations, we show that instability times obey a power-law scaling with initial planet spacing (measured in units ${\propto}\,\fracbrac{a}{\Delta a}\fracbrac{m}{M_*}^{1/4}$), with parameters that depend on the initial normalized eccentricities of the planets. Planet separations are therefore better reported in units that scale with mass according to ${\propto}\,\fracbrac{m}{M_*}^{1/4}$, rather than in units of mutual Hill radii (which scale with mass as ${\propto}\,\fracbrac{m}{M_*}^{1/3}$).

Based on our $N$-body results, we point out that multiplanet systems comprised of three or more sub-Neptune-mass planets must be spaced with period ratios $\mathcal{P}\,{\gtrsim}\,1.35$ to be stable over $10^9\,P_1$, in agreement with the spacings of the most compact observed multiplanet systems. Similarly, the strong dependence of instability time on eccentricity can be leveraged to place upper limits on the planets' eccentricities; tightly spaced observed systems ($\mathcal{P}\,{\lesssim}\,1.5$) must possess very low eccentricities ($e\,{\lesssim}\,0.05$) to be stable over $10^9\,P_1$. Constraints for different planet spacings/masses can be extracted from our empirical fit (Eq.~\ref{eq:linear_fit}), which accurately predicts instability times up to at least $10^9\,P_1$ for systems with a broad range of planet masses, spacings, and eccentricities.

This work represents a step toward understanding the dynamical processes that shape the population of observed multiplanet systems. The analysis presented in this paper was enabled by adopting a novel semi-analytic approach, made easy by the open-source \celmech code. We expect a similar approach to be useful for distilling the key dynamics of other problems in celestial mechanics.

\section{Acknowledgments} 
\label{sec:acknowledgments}

We thank the referee, Antoine Petit, for a thorough review and valuable comments. We also thank Dan Tamayo, Yubo Su, and Konstantin Batygin for useful discussions. Simulations were performed on the Sunnyvale cluster at the Canadian Institute for Theoretical Astrophysics. We acknowledge the support of the Natural Sciences and Engineering Research Council of Canada (NSERC; RGPIN-2023-04901). This work was performed in part at the Aspen Center for Physics, which is supported by National Science Foundation grant PHY-2210452. This research has made use of the NASA Exoplanet Archive, which is operated by the California Institute of Technology, under contract with the National Aeronautics and Space Administration under the Exoplanet Exploration Program.

\appendix

\section{Dynamical model integrator}
\label{sec:integrators}

\begin{figure}
\centering
\includegraphics[width=0.40\textwidth]{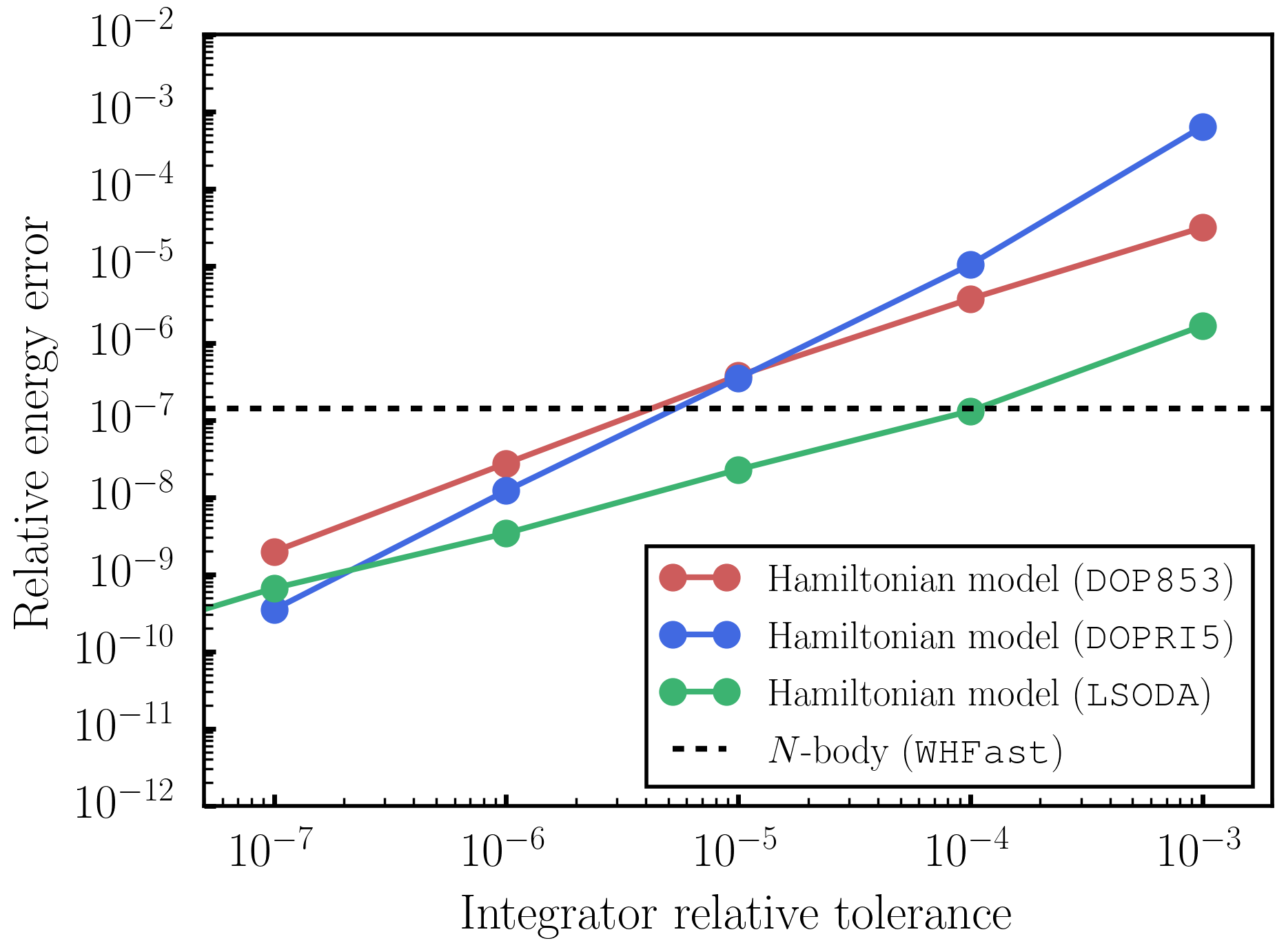}
\caption{Relative energy error of dynamical model integrations after $10^7\,P_1$ vs.\ the relative tolerance of the integrator (\texttt{DOP853}, \texttt{DOPRI5}, or \texttt{LSODA}) on an example system that is stable over $10^7\,P_1$. We use Model (2) to benchmark the integrators. With a relative tolerance of $10^{-5}$ for \texttt{DOP853} and \texttt{DOPRI5} and $10^{-4}$ for \texttt{LSODA}, the generic \texttt{SciPy} integrators achieve a similar energy error ($\sim\,10^{-7}$) to an $N$-body integration with \texttt{WHFast} using a time step of $P_1/20$.}
\label{fig:benchmark}
\end{figure}

We integrate the equations of motion generated by our dynamical models using the \texttt{SciPy} package's \texttt{integrate.ode} class, which offers several generic integrators \citep{Virtanen2020}. In contrast with the fixed-time-step, symplectic integrators often used for the full $N$-body problem (e.g., \texttt{WHFast}; \citealt{Rein2015}), these integrators dynamically adapt their time step to enforce absolute and relative tolerances on the local truncation errors. For simplicity, we fix the absolute tolerance to zero and choose the relative tolerance. To quantify the performance of the \texttt{SciPy} integrators on this problem, Figure~\ref{fig:benchmark} shows the energy error after integrating an example system for $10^7\,P_1$ with three generic integrators: \texttt{LSODA} \citep{Petzold1983}, and two Dormand-Prince integrators \citep{Dormand&Prince1978}, \texttt{DOPRI5} and \texttt{DOP853}. The benchmark system consists of five Earth-mass planets on initially circular orbits, with $\mathcal{P}\,{=}\,1.23$ and random orbital angles. We benchmark the integrators with Model (2), as shown in the middle panel of Fig.~\ref{fig:e0.0MMRs}. \texttt{DOP853} and \texttt{DOPRI5} achieve a similar energy error to \texttt{REBOUND}'s \texttt{WHFast} (${\sim}\,10^{-7}$) with a relative tolerance of $10^{-5}$ and \texttt{LSODA} does so with a relative tolerance of $10^{-4}$. With these tolerances, \texttt{DOP853} is faster than \texttt{LSODA} (by a factor of ${\sim}\,2$) and slightly faster than \texttt{DOPRI5} (by ${\sim}\,10$\%). As a result, we adopt \texttt{DOP853} with a relative tolerance of $10^{-5}$ for the integrations presented in this work. We have validated a subset of our dynamical model integrations with an expensive symplectic integrator and found no noticeable change in the recorded instability times. Note that, even with \texttt{DOP853}, integrating the equations of motion generated by a dynamical model that includes many resonances is slower than direct $N$-body integrations with \texttt{WHFast} by a factor of ${\sim}\,7$\,--\,$150$ depending on the complexity of the model.

Our finding that instability times can be accurately recovered by considering only a handful of resonances (e.g., Fig.~\ref{fig:e0.0MMRs}) suggests that it may be possible to speed up $N$-body integrations of multiplanet systems for some purposes. It may therefore be fruitful to develop optimized, symplectic integrators for simplified dynamical model integrations. An example illustrating how to set up and integrate dynamical models with \celmech can be found at \url{https://github.com/shadden/celmech/blob/master/jupyter_examples/Lammers24DynamicalModels.ipynb}.

\section{The location of first-order MMRs}
\label{sec:MMRloc}

\begin{figure}
\centering
\includegraphics[width=0.40\textwidth]{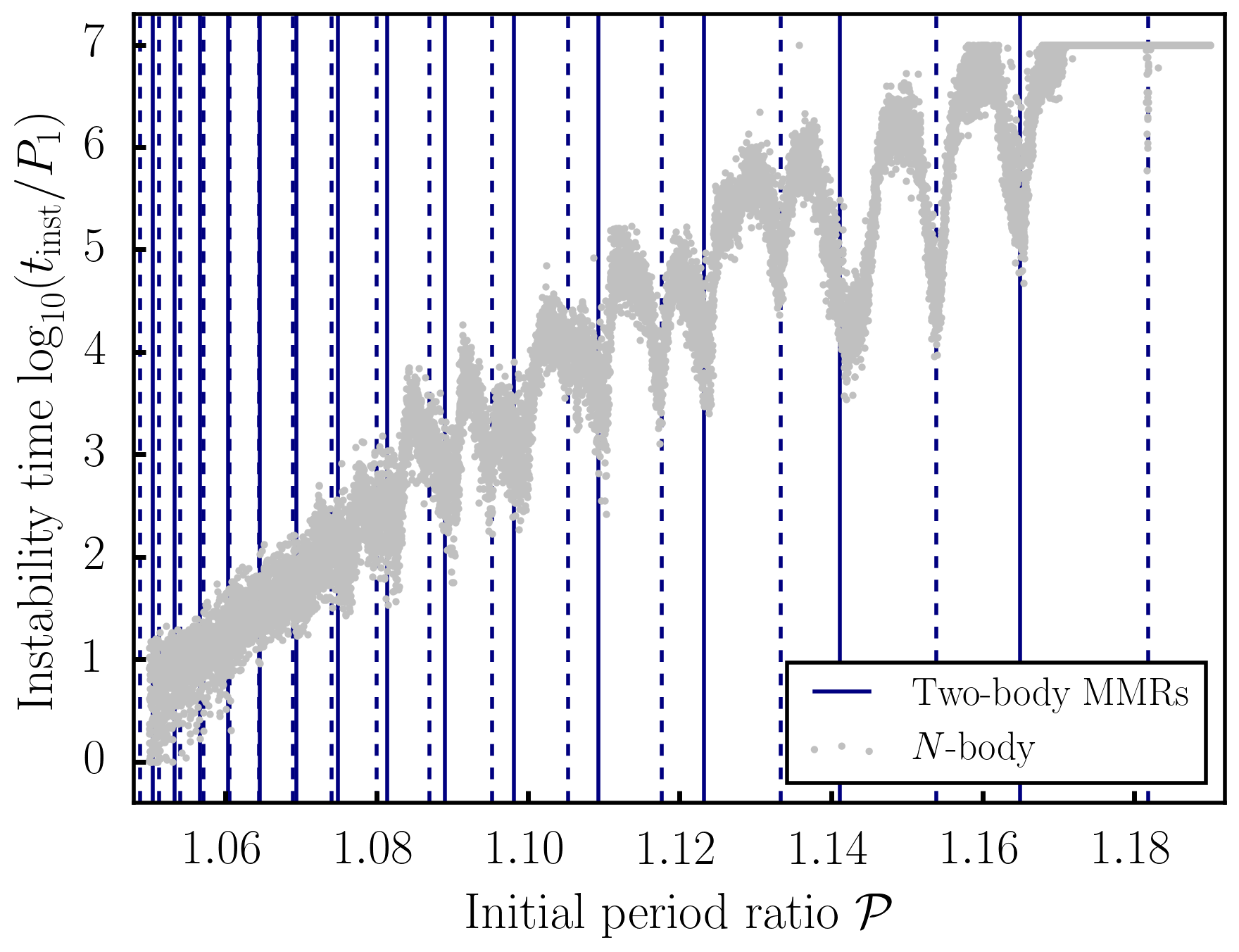}
\caption{Instability times for initially circular systems (left panel of Fig.~\ref{fig:e0.0MMRs}), with first-order MMRs placed at their true locations rather than $\mathcal{P}\,{=}\,p/(p-1)$. Dips in survival time occur slightly to the left of $\mathcal{P}\,{=}\,p/(p-1)$ (see Fig.~\ref{fig:e0.0MMRs}), lining up with the true location of first-order MMRs. The dip at $\mathcal{P}\,{\approx}\,1.143$ does not occur at the location of the 8:7 MMR because it is caused by the 3:2 MMR between planets $1-4$ and $2-5$ (see Fig.~\ref{fig:e0.0otherMMRs}).}
\label{fig:e0.0MMRloc}
\end{figure}

It was first noticed in \citet{Obertas2017} that first-order MMRs between adjacent planets lie at slightly larger period ratios than where dips in instability time occur (see Fig.~\ref{fig:e0.0MMRs}). In this appendix, we show that this offset is explained by the geometry of first-order MMRs. In particular, at zero eccentricity the separatrix of a first-order MMR lies at a slightly smaller period ratio than $\mathcal{P}\,{=}\,p/(p-1)$. We calculate this offset explicitly below and show that dips in survival time line up with the separatrix locations of first-order MMRs in Fig.~\ref{fig:e0.0MMRloc}.

The Hamiltonian governing the dynamics of a pair of planets in a first-order MMR is given, to lowest order in eccentricity, by
    \begin{equation}
       H = -\frac{1}{2}A(J-J^*)^2 - \tilde{\epsilon}\sqrt{J}\cos(\theta)
        \label{eq:first_order_mmr_ham}
    \end{equation}
(see Eq.~19 of \citealt{Hadden2019}). Introducing the canonical momentum-coordinate pair $(x,\,y)\,{=}\,\sqrt{2J}(\cos\theta,\,\sin\theta)$, the equations of motion derived from Hamiltonian \eqref{eq:first_order_mmr_ham} are given by
\begin{equation}
    \frac{d}{dt}\begin{pmatrix}x\\y\end{pmatrix}
    =
    \begin{pmatrix}
        A y \left(\frac{1}{2} \left(x^2+y^2\right)-J^*\right)\\
        -A x \left(\frac{1}{2} \left(x^2+y^2\right)-J^*\right)-\frac{\tilde{\epsilon}}{\sqrt{2}}
    \end{pmatrix}~.
\end{equation}
These equations have equilibria at $y\,{=}\,0$ and values of $x$ satisfying
\begin{equation}
    \frac{\sqrt{2} \tilde{\epsilon} }{A}-2 J^* x+x^3 = 0~.
    \label{eq:equilibrium_cond}
\end{equation}
Equation~\eqref{eq:equilibrium_cond} has either one or three real roots depending on the value of $J^*$. Specifically, there are three real roots when $J^*\,{>}\,\frac{3\tilde{\epsilon}^{2/3}}{2^{4/3} A^{2/3}}$. We follow \citet[][Appendix C]{FerrazMello2007} and 
use the identity $4\sin(x)^3\,{=}\,3\sin(x)\,{-}\,\sin(3x)$, to write the three equilibria as
\begin{equation}
    x_{\mathrm{eq},n} = \sqrt{\frac{8}{3}J^*} \sin \left(
    \frac{1}{3} \sin ^{-1}\left(\frac{3 \sqrt{3} \tilde{\epsilon}}{4 A (J^*)^{3/2}}\right)+\frac{2 \pi n}{3} \right) \mathrm{~;~} n=0,1,2
\end{equation}
where $n\,{=}\,0$ and $n\,{=}\,2$ are elliptic fixed points and $n\,{=}\,1$ is a hyperbolic fixed point. Let $x_\pm$ be the $x$-values of the upper and lower branches of the separatrix when $y\,{=}\,0$. Then we have that
\begin{equation}
    -\frac{8}{A}(H(x,0) - H(x_{\mathrm{eq},1},0)) = (x-x_{\mathrm{eq},1})^2(x-x_+)(x-x_-)~.
    \label{eq:sx_poly}
\end{equation}
Expanding the left-hand side of Eq.~\eqref{eq:sx_poly} shows that there is no cubic term, and thus $x_{\mathrm{eq},\,2}\,{=}\,-\frac{1}{2}(x_+\,{+}\,x_-)$. Equating the coefficients of the quadratic terms on the left and right of Eq.~\eqref{eq:sx_poly} gives $-4J^*\,{=}\,x_{\mathrm{eq},\,1}^2\,{+}\,2x_{\mathrm{eq},\,1}(x_-\,{+}\,x_+)\,{+}\,x_+x_-$. Together, these yield the solutions
\begin{equation}
x_\pm = -x_{\mathrm{eq},1} \pm 2\sqrt{ J^* - \frac{1}{2}x_{\mathrm{eq},1}^2}
\end{equation}
as the locations of the upper and lower separatrix branches. The upper separatrix branch sits at zero eccentricity (i.e., $x_+\,{=}\,0$) when $\frac{3}{4}x_\mathrm{eq}^2\,{=}\,J^*$, which occurs when $J^*\,{=}\,\frac{3}{2}\left(\frac{\tilde{\epsilon}}{A}\right)^{2/3}$. Using Eq.~21 of \citet{Hadden2019},
\begin{equation}
    \Delta \equiv \frac{j-1}{j} \frac{P_{2}}{P_1} - 1 = \frac{A(J-J^*)}{j}~,
\end{equation}
and so the separatrix crosses $e_1=e_2=0$ (for which $J = 0$) at
\begin{equation}
    \Delta_\mathrm{sx} = -\frac{3}{2j}A^{1/3}\tilde{\epsilon}^{2/3}
\end{equation}
where 
\begin{align}
    A &= \frac{3j(\mu_1+\mu_2)}{2}\left(\frac{j}{\mu_2} + \frac{j-1}{\mu_1\sqrt{\alpha}}\right)\\
    \tilde{\epsilon} &= 2\alpha^{-1/4}\left(\frac{\mu_1 \mu_2}{\mu_1+\mu_2}\right)^{1/2} {\sqrt{\mu_2 f^2 +\mu_1\sqrt{\alpha } g^2 }}~.
\end{align}
Here, $\mu_i$ is the star-to-planet mass of the $i$th planet, $\alpha\,{=}\,a_1/a_2\,{=}\,(P_1/P_2)^{2/3}$, and $f$ and $g$ are the disturbing function coefficients, $C_{(j,\,1\,{-}\,j,\,-1,\,0,\,0,\,0)}^{(0,\,0,\,0,\,0)}$ and $C_{(j,\,1\,{-}\,j,\,0,\,-1,\,0,\,0)}^{(0,\,0,\,0,\,0)}$, respectively, in the notation of \citet{celmech2022}.

\section{Three-body resonance widths}
\label{appendix:3br_widths}

Here, we provide a derivation of the widths of the 3BRs that are (implicitly) included in our simplified dynamical models. Let us begin by writing the interaction Hamiltonians between adjacent planets included in our dynamical models as the sum of the contributions from individual resonant terms. In order to do so, it will be convenient to first introduce the canonical action variables, $\Lambda_i\,{=}\,m_i\sqrt{GM_*a_i}$ conjugate to planets' mean longitudes, $\lambda_i$, along with the complex canonical variables,
\begin{eqnarray}
    x_i = \sqrt{m_i}(GM_*a_i)^{1/4}\paren{1 - \sqrt{1-e_i^2}}^{1/2}\exp(\mathrm{i}\varpi_i)~,
\end{eqnarray}
and their complex conjugates, $\bar{x}_i$. The Poisson bracket of two functions of phase-space variables, $f$ and $g$, can then be written in terms of these variables as
\begin{equation}
    [f,g] = \sum_{i}\paren{\pd{f}{\lambda_i}\pd{g}{\Lambda_i} - \pd{f}{{\Lambda}_i}\pd{g}{{\lambda}_i}}-\mathrm{i}\paren{\pd{f}{x_i}\pd{g}{\bar{x}_i} - \pd{f}{\bar{x}_i}\pd{g}{{x}_i}}~.
\end{equation}
Furthermore, for a given dynamical model and period ratio, $\mathcal{P}$, let us denote the set of MMR terms included as a set, $\mathcal{R}(\mathcal{P})$, with integer pairs $(j,\,k)$ included for each set of $j$:$j-k$ MMR terms in the model. For example, for a system with period ratio $\frac{j+1}{j}\,{<}\,\mathcal{P}\,{<}\,\frac{j}{j-1}$, Model (2) corresponds to $\mathcal{R}(\mathcal{P})\,{=}\,\{(j,\,1),\,(j\,{+}\,1,\,1)\}$. We can then write the complete Hamiltonian of a given dynamical model compactly as
\begin{eqnarray}
    H(\Lambda,\lambda,x,\bar{x}) = H_\mathrm{Kep}(\Lambda) +\epsilon H_\mathrm{pert}(\lambda,x,\bar{x})~, \label{eq:ham}
\end{eqnarray}
where we use the notation $\Lambda\,{=}\,(\Lambda_1,\,\Lambda_2,\,...)$ (and similar for $\lambda,\, x,\,\mathrm{and}~\bar{x}$) and also introduce $\epsilon$ as a bookkeeping parameter to identify terms of like power in planet-to-star mass ratios in the calculations below. The Hamiltonian in Eq.~\eqref{eq:ham} is separated into a Keplerian piece:
\begin{eqnarray}
    H_\mathrm{Kep}(\Lambda) = -\sum_{i=1}^{5}\frac{(GM_*)^2m_i^3}{2\Lambda_i^2}~,
    \end{eqnarray}
and a perturbation piece:
\begin{eqnarray}
    \epsilon H_\mathrm{pert}(\lambda,x,\bar{x}) &= \sum_{i=1}^{4}\sum_{(j,k)\in\mathcal{R}(\mathcal{P})}\paren{H_{(j,k)}^{(i,i+1)} + \bar{H}_{(j,k)}^{(i,i+1)}}~,
\end{eqnarray}
where 
\begin{equation}
    H_{(j,k)}^{(i,i+1)} = P_{j,k}(\bar{x}_i,\bar{x}_{i+1};\alpha_{i,i+1})\exp[i(j\lambda_{i+1}+(k-j)\lambda_{i})]~.
\end{equation}
$P_{j,k}(\bar{x}_i,\,\bar{x}_{i\,{+}\,1};\,\alpha_{i,i+1})$ is an homogeneous polynomial of degree $k$ in the variables $\bar{x}_i$ and $\bar{x}_{i\,{+}\,1}$, and $\bar{H}_{(j,\,k)}^{(i,\,i\,{+}\,1)}$ denotes the complex conjugate of ${H}_{(j,k)}^{(i,\,i\,{+}\,1)}$. The coefficients $P_{j,k}$ depend on planet pairs' semi-major axis ratios, $\alpha_{i,\,i\,{+}\,1}$, which are approximated as fixed in our \celmech dynamical models.

Next, we construct a near-identity canonical transformation that eliminates resonant terms from the transformed Hamiltonian to first order in planet masses. We construct this transformation by means of the Lie series method with generating function, $\chi$, so that the transformed Hamiltonian is given by
\begin{equation}
        H' = \exp[\epsilon\pounds_\chi]H = H_\mathrm{kep} + \epsilon ( [H_\mathrm{kep},\chi]+ H_1) +\epsilon^2 \left(\frac{1}{2}[[H_\mathrm{kep},\chi],\chi] + [H_1,\chi] \right) + \mathcal{O}(\epsilon^3)~.
\end{equation}
A straightforward calculation shows that setting 
\begin{eqnarray}
    \chi = \sum_{i=1}^{4}\sum_{(j,k)\in\mathcal{R}(\mathcal{P})}\paren{\chi_{(j,k)}^{(i,i+1)} + \bar{\chi}_{(j,k)}^{(i,i+1)}}~,
\end{eqnarray}
where
\begin{equation}
    \chi_{(j,k)}^{(i,i+1)} = -\frac{\mathrm{i}H_{(j,k)}^{(i,i+1)}}{jn_{i+1}(\Lambda_{i+1})+(k-j)n_{i}(\Lambda_{i})}
\end{equation}
and $n_i(\Lambda_i)\,{=}\,(GM_*)^2m_i^3/\Lambda_i^3$ yields $[H_\mathrm{kep},\,\chi]\,{+}\,H_1\,{=}\,0$. Consequently, the transformed Hamiltonian is given by
\begin{eqnarray}
        H' = \exp[\epsilon\pounds_\chi]H = H_\mathrm{kep}  +\frac{1}{2}\epsilon^2 [H_1,\chi] + \mathcal{O}(\epsilon^3)~.
\end{eqnarray}
Terms of second order in planet masses in the new, transformed Hamiltonian are given explicitly by
\begin{multline}
    \frac{\epsilon^2}{2}\sum_{i=1}^{4}\sum_{i'=1}^{4}\sum_{(j,k)\in\mathcal{R}(\mathcal{P})}
    \sum_{(j',k')\in\mathcal{R}(\mathcal{P})}\paren{
[H_{(j',k')}^{(i',i'+1)},\chi_{(j,k)}^{(i,i+1)}]
    +
[H_{(j',k')}^{(i',i'+1)} , \bar{\chi}_{(j,k)}^{(i,i+1)}]
+
[\bar{H}_{(j',k')}^{(i',i'+1)},\chi_{(j,k)}^{(i,i+1)} ]
+
[\bar{H}_{(j',k')}^{(i',i'+1)},\bar{\chi}_{(j,k)}^{(i,i+1)}]
    }~.\label{eq:full_sum_second_order}
\end{multline}
The Poisson brackets appearing in Eq.~\eqref{eq:full_sum_second_order} vanish unless $|i\,{-}\,i'|\,{\le}\,1$. Terms with $i\,{=}\,i'$ will produce new pairwise interaction terms in the transformed Hamiltonian. The effect of these terms on the dynamics will be negligible away from two-body MMRs. By contrast, terms with $i'\,{=}\,i\,{\pm}\,1$ produce new three-body interaction terms. Let us write these terms as $H_\mathrm{3BRs}\,{=}\,H_\mathrm{(3BRs,\,+)}\,{+}\,H_\mathrm{(3BRs,\,-)}$, where
\begin{eqnarray}
H_\mathrm{(3BRs,+)} = 
\frac{\epsilon^2}{2}\sum_{i=2}^{4}
\sum_{(j,k)\in\mathcal{R}(\mathcal{P})}
\sum_{(j',k')\in\mathcal{R}(\mathcal{P})}
\paren{
[H_{(j',k')}^{(i-1,i)},\chi_{(j,k)}^{(i,i+1)}]
+
[H_{(j,k)}^{(i,i+1)},\chi_{(j',k')}^{(i-1,i)}]
+ c.c.
}~,\nonumber\\
H_\mathrm{(3BRs,-)} = \frac{\epsilon^2}{2}\sum_{i=2}^{4}
\sum_{(j,k)\in\mathcal{R}(\mathcal{P})}
\sum_{(j',k')\in\mathcal{R}(\mathcal{P})}
\paren{
[H_{(j',k')}^{(i-1,i)} , \bar{\chi}_{(j,k)}^{(i,i+1)}]
+
[\bar{H}_{(j,k)}^{(i,i+1)} , {\chi}_{(j',k')}^{(i-1,i)}] + c.c.
}~,
\end{eqnarray}
where ``$c.c.$'' denotes the complex conjugate of the preceding term (note that $[\bar{f},\,\bar{g}]\,{=}\,\overline{[f,g]}$). Now, we define 3BR resonance amplitudes, ${Q}^{(i,\,\pm)}_{j',\,k',\,j,\,k}$, such that
\begin{eqnarray}
   [H_{(j',k')}^{(i-1,i)} , \bar{\chi}_{(j,k)}^{(i,i+1)}]
+
[\bar{H}_{(j,k)}^{(i,i+1)} , {\chi}_{(j',k')}^{(i-1,i)}] = {Q}^{(i,-)}_{j',k',j,k} \exp[-\mathrm{i}\paren{j\lambda_{i+1} + (k-j-j')\lambda_i + (j'-k')\lambda_{i-1}}]~,
\end{eqnarray}
\begin{eqnarray}
   [H_{(j',k')}^{(i-1,i)} , {\chi}_{(j,k)}^{(i,i+1)}]
+
[{H}_{(j,k)}^{(i,i+1)} , {\chi}_{(j',k')}^{(i-1,i)}] = {Q}^{(i,+)}_{j',k',j,k} \exp[\mathrm{i}\paren{j\lambda_{i+1} + (j+j'-k)\lambda_i + (k'-j')\lambda_{i-1}}]~.\label{eq:3br_res_amps_defn}
\end{eqnarray}
Evaluating the Poisson brackets and defining $\omega_{j,\,k}(n',\,n)\,{=}\,jn'\,{+}\,(k\,{-}\,j)n$, the amplitudes can be written as
\begin{eqnarray}
   {Q}^{(i,-)}_{j',k',j,k} &= 
    -
    \partial_{x_i}P_{j,k}({x}_{i},{x}_{i+1};\alpha_{i,i+1})
    \partial_{\bar{x}_i}P_{j',k'}(\bar{x}_{i-1},\bar{x}_i;\alpha_{i-1,i})
    \paren{
    \frac{\omega_{j',k'}(n_i,n_{i-1}) + \omega_{j,k}(n_{i+1},n_{i})}
    {\omega_{j',k'}(n_i,n_{i-1})\omega_{j,k}(n_{i+1},n_{i})} 
    }
    \nonumber\\
    &-
    j'(j-k)P_{j,k}({x}_{i},{x}_{i+1};\alpha_{i,i+1})
    P_{j',k'}(\bar{x}_{i-1},\bar{x}_i;\alpha_{i-1,i})
    \pd{n_i}{\Lambda_i}
    \paren{
    \frac{\omega_{j,k}(n_{i+1},n_{i})^2 + \omega_{j',k'}(n_{i+1},n_i)^2}
    {\omega_{j',k'}(n_{i+1},n_i)^2\omega_{j,k}(n_{i+1},n_{i})^2}
    }~,\label{eq:3br_amp_m}
\\
{Q}^{(i,+)}_{j',k',j,k} &= 
    j'(j-k)
    P_{j',k'}(\bar{x}_{i-1},\bar{x}_i;\alpha_{i-1,i})
    P_{j,k}(\bar{x}_{i},\bar{x}_{i+1};\alpha_{i,i+1})
    \pd{n_i}{\Lambda_i}
    \paren{
    \frac{\omega_{j,k}(n_{i+1},n_{i})^2 + \omega_{j',k'}(n_{i+1},n_i)^2}
    {\omega_{j',k'}(n_{i+1},n_i)^2\omega_{j,k}(n_{i+1},n_{i})^2}
    }~.\label{eq:3br_amp_p}
\end{eqnarray}
The expression for 3BR amplitudes given in Eqs.~\eqref{eq:3br_amp_m} and \eqref{eq:3br_amp_p} can be used to determine the extent of 3BRs in our dynamical models.

To simplify our notation, we will calculate the width of such a resonance between the first three planets in a system, noting that the calculation can be trivially generalized to any trio of adjacent planets. We consider the 3BRs involving a combination of mean longitudes $j_3\lambda_{3}\,{+}\,j_{2}\lambda_2\,{+}\,j_{1}\lambda_{1}$, where $(j_1,\,j_2,\,j_3)\,{=}\,(\pm(k'\,{-}\,j'),\,k\,{-}\,j\,{\pm}\,j',\,j)$. Let $\Lambda_{i,\,0}$ be the values of the canonical momenta $\Lambda_{i}$ at exact resonance, where $j_3n_{3}\,{+}\,j_{2}n_2\,{+}\,j_{1}n_{1}\,{=}\,0$, define $\delta\Lambda_i\,{=}\,\Lambda_{i}\,{-}\,\Lambda_{i,\,0}$, and write the Keplerian part of the Hamiltonian as
\begin{eqnarray}
    H_\mathrm{kep} \approx \sum_{i=1}^{3} n_{i,0}\delta\Lambda_i - \frac{3n_{i,0}}{2\Lambda_{i,0}}\delta\Lambda_i^2~.
\end{eqnarray}
We now define canonical coordinate variables $\theta\,{=}\,A\lambda$ where $A$ is any nonsingular matrix with $(A_{1,\,1},\,A_{1,\,2},\,A_{1,\,3})\,{=}\,(j_1,\,j_2,\,j_3)$. The conjugate momentum variables to these coordinate variables are given by $I\,{=}\,(A^\mathrm{T})^{-1}\delta\Lambda$. Ignoring nonresonant terms, the Hamiltonian governing the 3BR in terms of these new coordinates will be
\begin{eqnarray}
H' \approx H_\mathrm{kep} + H_\mathrm{(3BRs,\pm)} = -\frac{1}{2M_{j_1,j_2,j_3}} I_1^2 + \frac{1}{2}\paren{Q^{(2,\pm)}_{j'k',j,k}e^{\pm\mathrm{i}\theta_1} + \bar{Q}^{(2,\pm)}_{j'k',j,k}e^{\mp\mathrm{i}\theta_1}}~,
\end{eqnarray}
where
\begin{eqnarray}
    \frac{1}{M_{j_1,j_2,j_3}} = -\sum_{i=1}^{3} j_i^2\frac{3n_i}{\Lambda_{i,0}}~.
\end{eqnarray}
If the time evolution of the complex eccentricity variables, $x_i$, is ignored, then this is the Hamiltonian of a pendulum with maximal libration width given by
\begin{eqnarray}
    \Delta I_{1,\mathrm{max}} = 2\sqrt{|M_{j_1,j_2,j_3}Q^{(2,\pm)}_{j'k',j,k}|}~.
\end{eqnarray}
The corresponding maximal excursions in orbital frequencies are then given by
\begin{eqnarray}
    n_{i,\pm} = n_{i,0}\paren{1 \pm \frac{3j_i}{\Lambda_{i,0}} \Delta I_{1,\mathrm{max}}}~.\label{eq:3br_width_in_n}
\end{eqnarray}

Finally, we remark that the strengths of 3BRs arising in the actual planetary $N$-body problem will differ slightly from those computed for our dynamical models, not only because we have truncated our disturbing function expansion at lowest order in eccentricities, but also because the 3BR angles we consider can be formed via additional combinations of two-body resonances that are omitted from our dynamical models. For example, the $9\lambda_3\,{-}\,17\lambda_2\,{+}\,8\lambda_1$ three-body MMR plotted in Fig.~\ref{fig:mmr_web} is formed by the combination of the $9\lambda_3\,{-}\,8\lambda_2$ and $9\lambda_2\,{-}\,8\lambda_1$ first-order two-body MMR terms. This same angle combination can be created from the $9(\lambda_3\,{-}\,\lambda_2)$ and $8(\lambda_2\,{-}\,\lambda_1)$ zeroth-order two-body terms, the $9\lambda_3\,{-}\,7\lambda_2$ and $2(5\lambda_2\,{-}\,4\lambda_1)$, second-order two-body terms, and so on. Note that \citet{Petit2020} include the contributions arising from pairs of two-body zeroth-order MMRs when computing the strengths of zeroth-order 3BRs. However, we find that the magnitude of these contributions is almost always negligible, even at period spacings far from the pairs of first-order two-body MMRs that generate these 3BRs.

\bibliography{refs}{}
\bibliographystyle{aasjournal}

\end{document}